\documentclass[pre,longbibliography,twocolumn]{revtex4-2}
\usepackage[utf8]{inputenc}
\usepackage{natbib}
\usepackage{amssymb,amsmath}
\usepackage{epsfig}
\usepackage{bm}
\usepackage{soul}
\usepackage{xcolor}

\usepackage{graphicx}

\newcommand{\sbyr}[1]{{ #1}}

\newcommand\beq{\begin{equation}}
\newcommand\eeq{\end{equation}}
\newcommand\beqa{\begin{eqnarray}}
\newcommand\eeqa{\end{eqnarray}}

\newcommand{\rb}{\mathbf{r}}
\newcommand{\vb}{\mathbf{v}}
\newcommand{\yb}{\mathbf{y}}
\newcommand{\la}{\left\langle}
\newcommand{\ra}{\right\rangle}

\begin{document}

\title{Gaseous Diffusion as a Correlated Random Walk}

\author{Santos Bravo Yuste}
\email{santos@unex.es}
\homepage{https://fisteor.cms.unex.es/investigadores/\hspace{0pt}santos-bravo-yuste/}
\author{Rub\'en G\'omez Gonz\'alez}
\email{ruben@unex.es}
\author{Vicente Garz\'o}
\email{vicenteg@unex.es}
\homepage{\hspace{0pt}https://fisteor.cms.unex.es/investigadores/\hspace{0pt}vicente-garzo-puertos/
}
\affiliation{Departamento de F\'{\i}sica and Instituto de Computaci\'on Cient\'{\i}fica Avanzada (ICCAEx), Universidad de Extremadura, E-06006 Badajoz, Spain}

\begin{abstract}

\sbyr{The mean square displacement per collision of a molecule immersed in a gas at equilibrium is given by its mean square displacement between two consecutive collisions (mean square free path) corrected by a prefactor in the form of a series.
The $n$-th term of the series is proportional to the mean value of the scalar product $\rb_1 \cdot \rb_{n}$, where $\rb_i$ is the displacement of the molecule between the $(i-1)$-th and $i$-th collisions.
Simple arguments are used to obtain approximate expressions for each term. The key finding is that the ratio of consecutive terms in the series closely approximates the so-called mean persistence ratio. Exact expressions for the terms in the series are considered and their ratios for several consecutive terms are calculated for the case of hard spheres, showing an excellent agreement with the mean persistence ratio. These theoretical results are confirmed by  solving the Boltzmann equation by means of the direct simulation Monte Carlo method.
By summing the series, the mean square displacement and the diffusion coefficient can be determined using only two quantities: the mean square free path and the mean persistence ratio. A simple and an improved expression for the diffusion coefficient $D$ are considered and compared with the so-called first and second Sonine approximations to $D$ as well as with computer simulations of the Boltzmann equation. It is found that the improved diffusion coefficient shows very good agreement with simulation results over all intruder and molecule mass ranges.  When the intruder mass is smaller than that of the gas molecules, the improved formula even outperforms the first Sonine approximation.
}

\end{abstract}

\date{\today}
\maketitle


\section{Introduction}
\label{sec1}
The  Chapman--Enskog  solution of the Boltzmann equation is the most  successful and well-known  method for  determining the transport coefficients of molecular gases  \cite{CC70}.  This well-studied procedure  assumes the existence of a normal solution where all the space and time dependence of the velocity distribution function $f$ occurs only through a functional dependence on the hydrodynamic fields. This functional dependence can be made more explicit when the spatial gradients are small. In this situation, $f$ is written as a series expansion in powers of the spatial gradients of the hydrodynamic fields (density, mean flow velocity and temperature): $f=f^{(0)}+f^{(1)}+f^{(2)}+\cdots$. Here, $f^{(0)}$ is the local version of the Maxwell-Boltzmann distribution function (zeroth-order in gradients), $f^{(1)}$ is of first-order in gradients (Navier--Stokes order), $f^{(2)}$ is of second-order in gradients (Burnett order), \ldots.

Unfortunately,  within the Chapman--Enskog procedure, one has to be ready to pay the price of dealing with a mathematically complex and not very intuitive method.
On the other hand, since the beginning of the kinetic theory of gases with Maxwell, Stefan, Boltzmann, and Meyer \cite{C98}, there has been  a primary approach  to address the  study  of transport properties of gases, namely,  the mean free path (MFP) theory.  This theory  is much simpler and more intuitive than the Chapman--Enskog theory. This explains its use in  many  textbooks as an  introductory method for  studying  transport properties in gases  \cite{Reif1965,McQuarrie1976} and why is sometimes referred to as the \emph{elementary} kinetic theory of gases.

 In the elementary kinetic theory, the molecules are treated as masses of negligible volume (point particles) in constant random motion. They experience negligible intermolecular forces except during collisions, which justifies the concept of a free path between collisions.  However, this approach suffers from some shortcomings: (i) its results are not always accurate; (ii) its arguments are in many occasions qualitative and  highly debatable, and (iii) there is no general systematic way to improve its approximations \cite{Furry1948,CC70}.
Notwithstanding, numerous researchers have shown that many of these limitations can be lessened and that a free path approach can yield good predictions on the diffusion of gases while retaining its ability to offer insights into the underlying physical mechanisms \cite{Furry1948,Yang1949,Furry1951,Monchick1962,Kosov1982}.

In this paper we consider the problem of diffusion of an intruder in a dilute gas at equilibrium (intruder and particles of the gas are in general mechanically different, i.e., they can differ in mass and size).
In our approach, we treat the diffusion process of a molecule (or intruder) in the gas as a random walk problem.  In other words, we view this diffusion problem  as the result of a random process in which the molecule is, or behaves like, a random walker moving  with  steps  $\rb_i$  of random size and direction.  These steps (also referred to as flights) simply represent displacements of the molecule (or walker) between consecutive collisions. Alternatively, employing the classic terminology, they are recognized as the free paths.  In this paper we  calculate  the diffusion coefficient by evaluating the mean square displacement of the molecule (random walker). Authors such as Yang \cite{Yang1949}, Furry \cite{Furry1948}, Pitkanen \cite{Furry1951} and Monchick \cite{Monchick1962}  explored this avenue many years ago.
In particular, if one assumes that the steps $\rb_i$ are \textit{uncorrelated} (i.e., $\langle \rb_i\cdot\rb_{i+k}\rangle=0$, where $\langle \cdots \rangle$ denotes an average), then 
one finds a very simple expression for the diffusion coefficient given by \cite{Reif1965,McQuarrie1976}
\beq
\label{0.1}
D\equiv D_\text{eKT}=\frac{\lambda \bar{v}}{d},
\eeq
where $\lambda$ and $\bar{v}$ are the mean free path and the average velocity \sbyr{between two successive collisions},
respectively, of the molecule, and $d$ is the dimensionality of the system. In Eq.\ \eqref{0.1}, the subscript $\text{eKT}$ means that the diffusion coefficient has been evaluated within the elementary kinetic theory. Unfortunately, the expression \eqref{0.1}   is quite inaccurate.
This is primarily due to the assumption that the steps are uncorrelated, which is a  significant  oversimplification. On the contrary, when molecules collide they exhibit a strong tendency to continue in a direction similar to the one they had before the collision (i.e., $\langle \rb_i\cdot\rb_{i+k}\rangle>0$). This is the so-called  ``persistence'' of collisions, a fact employed in the past to improve the expression \eqref{0.1} \cite{Jeans1904,Jeans1940}.

Thus, for the reasons mentioned before, we will  consider in this paper the diffusion of an intruder in a gas at equilibrium as a random walk with \emph{correlated} steps (flights). We will show how the inclusion of these sort of correlations in an approximate and simple  way  yields  an expression for the diffusion coefficient $D$ that is fully consistent with the  classical  formula  derived by  Jeans \cite{Jeans1940}  taking into account the persistence.
We will subsequently extend our discussion and propose an improved expression of $D$. This new form of the diffusion coefficient turns out to be equivalent to an expression mentioned but not developed in the work of Jeans \cite{Jeans1940}.
It should be noted that the formulas  for the coefficient $D$ reported in this paper have been  derived in a different, independent and, in our opinion, more compelling approach  than the ones displayed in the work of Jeans \cite{Jeans1940}. This is one of the new added values of the present contribution. The key point in our  methodology  is the evaluation of the mean value of the projection of  $\rb_{i+k}$ over $\rb_{i}$, i.e,  $\langle \rb_i\cdot\rb_{i+k}\rangle$, which constitute the terms  appearing in the series that leads to the diffusion coefficient.
The evaluation of $\langle \rb_i\cdot\rb_{i+k}\rangle$ will allow us to estimate the de-correlation length;  it represents  the distance that the random walker has to travel to lose the memory of the size and direction of its initial step.
In this work,  we  focus on  a hard sphere gas as the specific system to apply our  methodology.   However, any other type of interaction could have been considered as long as the conditions for a free path description are met.

It is also interesting to note that, as shown by Monchik and Mason \cite{Monchick1962,Monchick1967}, there  exists  a very close relationship between the  results derived from  the Chapman--Enskog and MFP theories. It turns out that an integral equation for the velocity distribution function  \emph{coincides}  with the integral equation
 verifying the Chapman--Enskog approximation $f^{(1)}$   when considering the limit of infinitely many collisions in the MFP theory.  However, Monchik and Mason \cite{Monchick1967}  pointed out  that the velocity distributions corresponding to the two theories differ \emph{slightly} because of the use of slightly different auxiliary conditions.

The plan of the paper is as follows. In Sec.~\ref{sec:2}, we present the diffusion of an  intruder in a dilute gas at equilibrium  as a random walk  problem. We calculate  the mean square displacement and the diffusion coefficient as an infinite series (the  collisional series) with terms of the form  $\la \rb_1\cdot\rb_{k}\ra$.
In Sec.~\ref{sec:3}, we make some simple approximations to estimate the value of  each term  $\la \rb_1\cdot\rb_{k}\ra$
 in this series.  We then sum  the series and obtain the formula proposed by Jeans \cite{Jeans1940} for the diffusion coefficient $D$.
The expression of $D$  is improved in Sec.~\ref{sec:4} by incorporating  the exact value of  the mean square free path   $\la r_1^2\ra$. The general formulas for evaluating the averages $\la \rb_1\cdot\rb_{k}\ra$ in terms of the collision frequency and the so-called transition rate are developed  in Sec.~\ref{sec:5}.
We use these expressions to compute the first few terms of the collisional series for hard spheres. Notably, we find that  the  quotient of the successive terms of this series is very well approximated by the mean persistence ratio.  To gauge the accuracy of our expressions for $D$, we compare \sbyr{in Sec.\ \ref{sec:compa} this random walk expression  with the theoretical forms of $D$} obtained from the Chapman--Enskog method in the so-called first- and second-Sonine approximations and with the results obtained by numerically solving the Boltzmann equation by means of the direct simulation Monte Carlo (DSMC) method \cite{B94}.
We end the paper in  Sec.~\ref{sec:Conclu} with some conclusions and remarks.

\section{The molecule as a random walker. Mean square displacement}
\label{sec:2}

We consider a gas in equilibrium with no external fields.
\sbyr{The velocity distribution of the molecules is  then a Maxwell-Boltzmann distribution.}
We single out a molecule that can be either identical to or distinct from the other molecules.
We refer to this molecule as the random walker or intruder (a term commonly used in kinetic theory of  granular gases \cite{GarzoBook19}).  We are interested in understanding how far this molecule will be from its initial position $\mathbf{R}(t)$   after a given time $t$.  With no external field, the average of this quantity is
equal to zero:  $\langle \mathbf{R}(t)\rangle=0$. However, the second moment $\langle R^2(t)\rangle$, or mean square displacement (MSD), is  the relevant quantity because provides information about the extent to which these intruders or random walkers spread (diffuse) within the gas. In particular, the diffusion coefficient $D$ of the molecule in a  three-dimensional gas can be obtained from the long-time (i.e., very large number of collisions) value  of the quotient $\langle R^2(t)\rangle/6t$, namely,
\begin{equation}
\label{DfromR2}
D=\lim_{t\to \infty}\frac{\langle R^2(t)\rangle}{6t}.
\end{equation}
\sbyr{The diffusion of a molecule in a gas at equilibrium is normal so that the limit of Eq.~\eqref{DfromR2} exists.}

\subsection{Evaluation of  $\langle {R}^2_N \rangle$}

To evaluate $D$ or $\langle R^2(t)\rangle$, it is convenient to first consider the displacement of the random walker (the intruder) after $N$ steps (collisions):
\beq
\label{2.1}
\mathbf{R}_N= \sum_{i=1}^N  \rb_i,
\eeq
where $\rb_i$ is $i$-th displacement (flight, free path) of the molecule. From Eq.\ \eqref{2.1},  one finds
\begin{align}
\label{MSD2}
 {R}^2_N
  &=
   \sum_{i=1}^N   \rb_i \cdot \rb_i
  +  \sum_{i\neq j}^N   \rb_i  \cdot \rb_j   \\
  &=
   \sum_{i=1}^N  \rb_i  \cdot \rb_i
  +  \, 2 \sum_{i=1}^{N-1}   \rb_i  \cdot \rb_{i+1}
  +  \, 2 \sum_{i=1}^{N-2}   \rb_i  \cdot \rb_{i+2}  +\cdots
\end{align}
At equilibrium, the probability density function (pdf) of the step sizes does not depend on time, so that
\beq
\label{MSDles2}
 \langle {R}^2_N\rangle
 =
N \langle \rb_1\cdot \rb_{1} \rangle  +2(N-1)   \langle \rb_1\cdot \rb_{2}\rangle
+2(N-2) \langle \rb_1\cdot \rb_{3}\rangle+\cdots
\eeq
\sbyr{It is convenient to} write $\langle {R}^2_N\rangle$ as
\begin{equation}
\label{MSDle3}
\langle {R}^2_N\rangle=N\ell_e^2,
\end{equation}
where $\ell_e$ can be identified as an \emph{effective} free path. The quantity \sbyr{$\ell_e^2=\langle {R}^2_N\rangle/N$, is the MSD per collision and} can be approximated by
\beq
\label{ell2Sume}
\ell_e^2\approx \langle  r_1^2  \rangle +  2\langle \rb_1\cdot \rb_{2}\rangle + 2\langle \rb_1\cdot \rb_{3}\rangle +\cdots
\eeq
for sufficiently large $N$. Here, we have assumed that the steps decorrelate fast enough so that the error in the approximation
\begin{equation}
\label{apdefast}
 (N-k-1) \,\langle \rb_1\cdot \rb_{k}\rangle \approx N \langle \rb_1\cdot \rb_{k}\rangle
\end{equation}
 can be ignored even for large  $k$.
For example, we will see in Sec.~\ref{sec:3} that, for elastic hard spheres where the mass of the intruder  $m_1$  is equal to the mass $m_2$ of the molecules of the gas,
the quantity $\langle \rb_1\cdot \rb_{1+k}\rangle$ decays exponentially fast, approximately as  $0.406^{k}$.
The validity of this approximation will be discussed in  Sec.~\ref{sec:compa}.

It is convenient to rewrite Eq.~\eqref{ell2Sume} as
\beq
\label{ell2redSer}
\frac{\ell_e^2}{\langle  r_1^2  \rangle}\approx   1+ \frac{2\langle \rb_1\cdot \rb_{2}\rangle}{\langle  r_1^2  \rangle}+   \frac{2\langle \rb_1\cdot \rb_{3}\rangle}{\langle  r_1^2  \rangle}+\cdots \equiv \sum_{n=1}^\infty c_n.
\eeq
We will call the series in Eq.~\eqref{ell2Sume} the \emph{collisional series} while the series $\sum_{n=1}^\infty c_n$ appearing in Eq.\ \eqref{ell2redSer} will be referred to as the \emph{reduced} collisional series.  In summary, Eqs.~\eqref{MSDle3} and \eqref{ell2redSer} tell us that the mean square displacement of the intruder per \sbyr{collision}, $\langle {R}^2_N\rangle/N$, is just the mean square free path, $\langle  r_1^2  \rangle$, corrected by the reduced collisional series.

The expression \eqref{MSDle3} for the MSD is exact for a  random walk  with  \emph{isotropic}  and \emph{uncorrelated} steps of a fixed size $\ell_e$.  This effective random walk [where  $\ell_e$  is determined by Eq.~\eqref{ell2Sume}] is  then \emph{equivalent} to the actual random walk in the sense that both lead to the same MSD.

\subsection{Evaluation of $\langle R^2(t) \rangle $ from  $\langle R^2_N \rangle$}

The pdf of the position  $\mathbf{R}$ of the intruder at time $t$ can be written as
\begin{equation}
\label{x0}
P(\mathbf{R},t)=\sum_{N=0}^\infty P_N(\mathbf{R}) \chi_N(t)
\end{equation}
where $P_N(\mathbf{R})$  is the  pdf of the the position of the walker after $N$ steps and  $\chi_N(t)$ is the probability of taking exactly $N$ steps up to time $t$. In terms of $P(\mathbf{R},t)$, the MSD is
\beq
\label{2.2}
\langle R^2(t) \rangle
=\int d\mathbf{R}\, P(\mathbf{R},t) R^2.
\eeq
Substitution of Eq.\ \eqref{x0} into Eq.\ \eqref{2.2} leads to
\begin{align}
\label{xl}
\langle R^2(t) \rangle=\sum_{N=0}^\infty  \chi_N(t) \int   P_N(\rb) R^2\, d\mathbf{R}
=\sum_{N=0}^\infty  \chi_N(t) \langle R^2_N \rangle.
\end{align}
On obtaining Eq.\ \eqref{xl} summation and integration have been interchanged.
Taking into account Eq.~\eqref{MSDle3},  we find
\begin{equation}
\label{MSDelle}
\langle R^2(t) \rangle
=\ell_e^2\sum_{N=0}^\infty  \chi_N(t) N =\langle N(t)\rangle \,\ell_e^2,
\end{equation}
$\langle N(t)\rangle$ being the average number of steps given (collisions suffered) by the random walker up to time $t$. From  Eqs.~\eqref{DfromR2} and \eqref{MSDelle}, one achieves the result
\begin{equation}
\label{DfromN}
D=\lim_{t\to \infty}\frac{\langle N(t)\rangle }{6t}\,\ell_e^2.
\end{equation}
 When the gas is at equilibrium, the collision frequency does not depend of the position and time. Therefore, $\langle N(t)\rangle /t$ represents  the average collision frequency $\langle \mu \rangle$ (recall that it is assumed that $t$ is large in this expression, implying a significant number of collisions).  Because the average time between collisions $\la \tau \ra$ (collision interval) is the inverse of the collision frequency \sbyr{\cite{Paik2014,Reif1965,Yang1949},}
then Eq.\ \eqref{DfromN} can be written as
\begin{equation}
\label{DfromNbis}
D=\frac{\langle \mu \rangle\,\ell_e^2 }{6}=\frac{\ell_e^2 }{6\langle \tau   \rangle}.
\end{equation}

It is convenient to rewrite Eq.\ \eqref{DfromNbis} in terms of the mean free path $\lambda$ and  the elementary kinetic theory diffusion coefficient $D_{\text{eKT}}$:
\begin{equation}
\label{Delle}
D=D_{\text{eKT}}\, \frac{ \ell_e^2}{2\lambda^2}.
\end{equation}
On writing Eq.\ \eqref{Delle}, use has been made of Eq.\ \eqref{0.1} and the fact  that the average speed $\bar{v}$ of the molecules is $\bar{v}=\lambda/\la \tau \ra=\lambda \langle \mu\rangle$ \cite{Paik2014,Reif1965,Yang1949}.
Using Eq.~\eqref{ell2Sume}, Eq.\ \eqref{Delle} can be finally written as
\begin{equation}
\label{DelleD0Ser}
D=D_{\text{eKT}}\,\frac{ \langle r_1^2\rangle }{2\lambda^2} \left(1+ \frac{2\langle \rb_1\cdot \rb_{2}\rangle}{\langle  r_1^2  \rangle}+   \frac{2\langle \rb_1\cdot \rb_{3}\rangle}{\langle  r_1^2  \rangle}+\cdots\right).
\end{equation}

 Assuming that $\lambda$ and $\langle \mu \rangle$ are known, the only quantity to evaluate for estimating $D$ is  the effective mean free path $\ell_e^2$. This is tantamount  to evaluate the averages $\langle \mathbf{r}_1 \cdot \mathbf{r}_{k} \rangle$ [see Eq.\ \eqref{ell2Sume}]. Since this task is not simple (as discussed later in Sec.\ \ref{sec:5}), we can make  some  simple approximations. Firstly, we take $\langle r_1^2 \rangle \approx 2\lambda^2$, which would be true if the gas molecules (except the intruder) were  motionless   \cite{Reif1965,McQuarrie1976,Paik2014}. Secondly, we neglect correlations between successive steps (i.e., $\langle \mathbf{r}_1 \cdot \mathbf{r}_{k} \rangle = 0$ for $k > 0$).
 Combining both approximations, we find that $\ell_e^2 = 2\lambda^2$ and so, the diffusion coefficient is given by the elementary kinetic theory \cite{Furry1951}, Eq.\ \eqref{0.1}.

\section{Correlations, persistence and MSD. Jeans formula}
\label{sec:3}

 In the previous Section \ref{sec:2}, we have discussed how  the evaluation of the intruder's MSD (or equivalently, its diffusion coefficient) can be reduced to the evaluation of the effective free path $\ell_e$ [see Eq.~\eqref{Delle}], or the evaluation of $\langle \rb_1\cdot \rb_{k}\rangle$ [see Eq.~\eqref{ell2Sume}].   The precise computation of these quantities will be addressed in  Sec.~\ref{sec:5}. In the present Section we show how to approximate them in terms of the mean free path $\lambda$ and the so-called persistence of the collisions. We employ relatively straightforward and simple arguments.

The evaluation of $\langle \rb_1\cdot \rb_{k}\rangle$ for $k=1$ (that is, $\langle r_1^2\rangle$) is the easiest one.  For elastic hard spheres,  an  elementary computation yields $\langle r_1^2\rangle\approx 2\lambda^2$. This comes from the simple (approximate) expression $P(\ell)=\exp(-\ell/\lambda)/\lambda$ for the  distribution of the free path length ($\ell\equiv r_1$) \cite{McQuarrie1976,Paik2014}.  This expression  is derived under the simplifying assumption that the intruder moves within a gas  where its particles are ``frozen''.
The rigorous result is more complex and we will discuss this in Sec.~\ref{sec:r12}.

For estimating the average
$\langle \rb_1\cdot \rb_{k}\rangle=\langle  {r}_1  {r}_{k} \cos \theta_{1,k}\rangle$ with $k\ge 2$
 we  assume  that the size $r_1$ of the step 1 and the value of the  projection of  the step $\rb_k$ over $\rb_1$ are uncorrelated.  Namely, we make  the  approximation
  \begin{equation}
  \label{r1rkA}
\langle \rb_1\cdot \rb_{k}\rangle\approx \langle  {r}_1\rangle  \langle {r}_{k}  \cos \theta_{1,k}\rangle=\lambda  \langle {r}_{k} \cos \theta_{1,k}\rangle,
  \end{equation}
where $\theta_{i,j}$  denotes the angle between the displacements $\rb_i$ and $\rb_j$.  It is important to note that correlations indeed exist due to the fact that the size of displacements (free paths) before and after the collision ($\rb_i$ and $\rb_{i+1}$, respectively) as well as the angle formed by these displacements depend on the corresponding velocities $\mathbf{v}_i$ and $\mathbf{v}_{i+1}$, which are correlated.

Let's now see how to estimate the term $\langle {r}_{k} \cos \theta_{1,k}\rangle$ by  using the spherical cosine law that relates the cosines of the angles $\theta_{1,k}$,  $\theta_{1,2}$ and $\theta_{2,k}$:
\begin{equation}
\label{cos1k}
\cos\theta_{1,k}=\cos\theta_{1,2}\,\cos\theta_{2,k} + \sin\theta_{1,2}\sin\theta_{2,k} \cos\varphi_{1,2,k},
\end{equation}
where  $\varphi_{1,2,k}$ is the angle between the plane  generated by $\{ \rb_1,\rb_2 \}$  and the plane generated by $\{ \rb_{2},\rb_{k} \}$.
Due to the rotational symmetry of collisions along the direction of the precollisional displacements   (or velocities), all values of  $\varphi_{1,2,k}$ are equally probable.  This implies that $\langle {r}_{k} \sin\theta_{1,2}\sin\theta_{2,k} \cos\varphi_{1,2,k} \rangle=0$, and then
 \begin{equation}
\label{x2}
\langle {r}_{k} \cos\theta_{1k}\rangle=\langle {r}_{k}\cos\theta_{1,2} \cos\theta_{2,k}\rangle.
\end{equation}
The relation \eqref{x2} can be rewritten as
\begin{equation}
\label{rkfac1}
\langle {r}_{k} \cos\theta_{1k}\rangle= \la  r_1\, \frac{r_2}{r_1} \cos\theta_{1,2} \frac{r_k}{r_2}\cos\theta_{2,k}\ra
\end{equation}
or, neglecting correlations, as
\begin{equation}
\label{rkfac2}
\langle {r}_{k} \cos\theta_{1k}\rangle
\approx
\lambda  \la\frac{r_2}{r_1} \cos\theta_{1,2}\ra \la\frac{r_k}{r_2} \cos\theta_{2,k}\ra.
\end{equation}
We repeat this  procedure for estimating the value of the last factor  $\la (r_k/r_2)\cos \theta_{2,k}\ra$. We use again the spherical cosine law that relates $\theta_{2,k}$ with $\theta_{2,3}$ and $\theta_{3,k}$.  As before, $\la ({r}_{k}/r_2) \sin\theta_{2,3}\sin\theta_{k-1,k} \cos\varphi_{2,3,k}\ra=0$ and so,
\begin{align}
\la\frac{r_k}{r_2} \cos\theta_{2,k}\ra=&
\la \frac{r_3}{r_2} \cos\theta_{2,3}\, \frac{r_k}{r_3} \cos\theta_{3,k}\ra  \nonumber\\
\approx& \la \frac{r_3}{r_2} \cos\theta_{2,3}\ra \, \la\frac{r_k}{r_3} \cos\theta_{3,k}\ra.
\label{rkfac3}
\end{align}
Repeating this procedure we get
\beqa
\label{rkfac}
\langle {r}_{k} \cos\theta_{1k}\rangle
&\approx&
\lambda \la\frac{r_2}{r_1} \cos\theta_{1,2}\ra \la\frac{r_3}{r_2} \cos\theta_{2,3}\ra
\cdots\nonumber\\
&=&
\lambda\,   \prod_{i=1}^{k-1}   \la\frac{r_{i+1}}{r_i} \cos\theta_{i,i+1}\ra.
\eeqa
On the other hand, at equilibrium, the averages of the form $\langle(r_{i+1}/r_i) \cos\theta_{i,i+1}\rangle$ do not depend on $i$ (i.e., on time), and then Eq.\ \eqref{rkfac} yields
\begin{equation}
\label{rcprod}
\langle {r}_{k} \cos\theta_{1k}\rangle
\approx
\lambda\,  \la\frac{r_2}{r_1} \cos\theta_{1,2}\ra^{k-1}.
\end{equation}
Following this, we approximate
\begin{equation}
\label{rrome}
\la\frac{r_2}{r_1} \cos\theta_{1,2}\ra \approx \la\frac{v_2}{v_1} \cos\theta_{1,2}\ra.
\end{equation}
Note that this relation would be exact if all the flight times (times between collisions) $\tau_i$  were the same for all $i$ as  $r_i=v_i \tau_i$.
Recall that $r_2 \cos\theta_{1,2}$ is the component of the postcollisional displacement $\rb_2$ along the direction of the precollisional displacement $\rb_1$. One could then see  $r_2 \cos\theta_{1,2}/r_1$ as the fraction of the precollisional displacement that is not lost, that persists, after the collision.  Similarly,  $v_2 \cos\theta_{1,2}/v_1$ is the fraction of $\vb_1$ that persists after a collision. The average of this quantity is the so-called mean persistence ratio $\omega$:
\beq
\label{w}
\la \frac{v_2}{v_1} \cos\theta_{1,2}\ra \equiv \omega.
\eeq
The mean persistence $\omega$ is a well-studied quantity in the kinetic theory of gases (see, for example,  section 5.5 of \cite{CC70}).  This explains our interest in  constructing  the factorizations of Eqs.~\eqref{rkfac1}, \eqref{rkfac3} and \eqref{rkfac}.

Collecting Eqs.~\eqref{r1rkA}, \eqref{rcprod}, \eqref{rrome}  and \eqref{rrome}, we finally find:
\begin{equation}
\label{rbrkok}
\la  \rb_1\cdot \rb_{k}\ra \approx \lambda^2\, \omega^{k-1}.
\end{equation}
An alternative deduction of this formula is presented in Appendix \ref{ap:alter}.
Equation~\eqref{rbrkok} together with   $\langle r_1^2\rangle\approx 2\lambda^2$ and Eq.~\eqref{ell2Sume} leads to
\beqa
\label{ele2pa}
\ell_e^2&\approx& 2\lambda^2\left(1+\omega+\omega^2+\cdots\right)
\nonumber\\
&=&\frac{2\lambda^2}{1-\omega},
\eeqa
which implies [see Eq.~\eqref{MSDelle}]
\begin{equation}
\label{R2ome}
\la R^2(t)\ra= \la N(t)\ra \,\frac{2 \lambda^2}{1-\omega}
\end{equation}
or [see Eq.~\eqref{Delle}]
\begin{equation}
\label{DJ}
D=\frac{D_{\text{eKT}}}{1-\omega}=\frac{\lambda \bar{v}}{3(1-\omega)} \equiv D_\text{J}.
\end{equation}
This simple and nice expression $D_\text{J}$ is no more than the classic  expression derived  by  Jeans (see the second formula of section 169 of Ref.\ \cite{Jeans1940}), which takes  into account the persistence of the molecules  to improve  the elementary kinetic expression $D_{\text{eKT}}$.

A full assessment of these formulas for elastic hard spheres will be given in Sec.~\ref{sec:compa}  when we compare them against DSMC simulations.  Nonetheless, we can  provide here  some numbers that  show  the large improvement that the consideration of the  persistence of the molecules leads to. For elastic hard spheres, one has $\omega\simeq 0.406$ for $m_1/m_2=1$,  $\omega\simeq 0.245$ for $m_1/m_2=0.5$, and $\omega\simeq 0.785$ for $m_1/m_2=5$.  Here, $m_1$ and $m_2$ are the masses of the intruder and the gas, respectively. From Eq.~\eqref{DJ}, one finds that $D/D_{\text{eKT}}$ is $1.68$ for $m_1/m_2=1$, $1.32$ for $m_1/m_2=0.5$, and  $4.66$ for $m_1/m_2=5$. The corresponding   DSMC simulation values  for $D/D_{\text{eKT}}$  are 1.40,  1.82, and  5.37, respectively.  The inclusion of correlations between displacements through the mean persistence ratio clearly improves the $D_{\text{eKT}}$ values.   More details will be given in Sec.~\ref{sec:compa}.

A long time ago it was realized (see Ref.\ \cite{Jeans1940}, for example) that one way to improve the accuracy of the MFP theory predictions was to take into account that, after a collision, the exit direction of the colliding molecule is not completely isotropic.  Instead, there  is a certain probability that the exit direction is similar to the direction of incidence. This is known as the persistence of velocities after collision \cite{Jeans1940,CC70}.   MFP-type formulas incorporating this aspect offer better and more accurate predictions. However,  the arguments used  to incorporate this fact into the equations were not entirely convincing. Perhaps for this reason, the consideration of arguments based on the persistence of collisions in the MFP formulation has not  gained  much success or relevance,  remaining  limited essentially to the contributions of Jeans \cite{Jeans1904,Jeans1940}. This is despite the clear and intuitive  understanding  that persistence must play a role in the transport properties, since these properties necessarily depend on the manner in which collisions occur. In the  present  Section, we have shown  how to include the persistence in the calculation of the diffusion coefficient using a different approach.  We will  further refine   this random walk approach in the subsequent Sections.

\section{An improved formula for the diffusion coefficient}
\label{sec:4}

One of the limitations of free path theory is its inability to control the reliability of its assumptions and approximations. This issue has been frequently highlighted, particularly  in the book of  Chapman  and Cowling  (see section 6.5 of Ref.~\cite{CC70}). However, it is important to recognize that the standard approach advocated by  the Chapman--Enskog method \cite{CC70} is not entirely immune to this limitation, as demonstrated by the widespread use of DSMC and molecular dynamics simulations  to ensure the results of the kinetic theory.

In this Section, we examine the assumptions underlying  Eqs.~\eqref{ele2pa} and \eqref{DJ}. Equation~\eqref{ele2pa} was obtained from  Eq.~\eqref{rbrkok} by assuming two key approximations: first,   $\la r_1^2\ra\approx 2\lambda^2$ and,  second, $\la  \rb_1\cdot \rb_{k}\ra \approx \lambda^2\, \omega^{k-1}$ for $k\ge 2$ [see  Eq.~\eqref{rbrkok}]
We can (partially) eliminate the first approximation by using the exact value of the mean square free path  $\la r_1^2\ra$.  In this case, from Eqs.\ \eqref{ell2Sume} and \eqref{rbrkok}, one has
\beqa
\label{ele2p}
 \ell_e^2&\approx& \la r_1^2\ra+2 \lambda^2\left(\omega+
 \omega^2+\cdots\right)\nonumber\\
 &=&
\la r_1^2\ra\left[1+\frac{2\lambda^2}{\la r_1^2\ra}\left(\omega+\omega^2+\cdots\right)\right]
\nonumber\\
&\approx&
 \la r_1^2\ra \left(1+\omega+\omega^2+\cdots\right)\nonumber\\
 &=&
 \frac{\la r_1^2\ra}{1-\omega},
\eeqa
where in the third line of Eq.\ \eqref{ele2p} we have replaced $2\lambda^2$ by $\la r_1^2\ra$. Substitution of Eq.\ \eqref{ele2p} into Eq.~\eqref{MSDelle} yields
\begin{equation}
\label{R2omep}\
\la R^2(t)\ra= \la N(t)\ra \,\frac{\la r_1^2\ra}{1-\omega}
\end{equation}
or, from Eq.~\eqref{DfromNbis}, one gets
\begin{equation}
\label{DR}
D=  \frac{\la r_1^2\ra}{6\la \tau \ra}\, \frac{1}{1-\omega}.
\end{equation}
It is convenient to rewrite this equation in terms of $D_\text{eKT}$ as
\begin{equation}
\label{DRW}
D= D_{\text{eKT}} \frac{\kappa}{1-\omega}\equiv D_\text{RW},
\end{equation}
where
\begin{equation}
\label{kappadef}
\kappa=\frac{\la r_1^2\ra}{2\lambda^2}.
\end{equation}
On writing Eq.\ \eqref{DRW}, use has been made of the Eq.~\eqref{0.1} and the relation $\bar{v}=\lambda/\la \tau \ra$.

The nice thing about Eqs.~\eqref{R2omep} and \eqref{DRW}  is that they are \emph{equivalent} to the formula
\begin{equation}
\label{DelleD0pJ}
D= \frac{\la v_1^2\tau_1\ra }{d } \frac{1}{1-\omega},
\end{equation}
which is obtained from Eq.~\eqref{DJ} if one replaces  in this equation $\lambda \bar{v}$ by $\la r_1  v_1\ra=\la v_1^2\tau_1\ra$. This replacement, mentioned by Jeans in Ref.~\cite{Jeans1940} (see the last equation of page 203 of this reference),  was not explored further, leaving its implications uncharted. Equations \eqref{DRW} and \eqref{DelleD0pJ} are equivalent provided that  $\langle r_1^2\rangle/\la \tau\ra=2 \la v_1^2\tau_1\ra$. This is proved in Sec.~\ref{sec:r12} [see Eq.~\eqref{r12MJ}].

It is important to note that the validity of Eq.~\eqref{ele2p}  [and equivalently,  Eqs.~\eqref{R2omep} and \eqref{DRW}], rests in the validity of the approximation
\begin{equation}
\label{sup2}
c_{1+k}=\frac{2 \la\rb_1\cdot \rb_{1+k}\ra}{\la r_1^2\ra} \approx \omega^{k},\quad \sbyr{k\ge 1}.
\end{equation}
As  mentioned  in Sec.~\ref{sec:2},  this result implies that  Eq.~\eqref{apdefast} holds unless $m_1/m_2\to\infty$ because (see at the end of Sec.~\ref{ap:basic}) $\omega\to 1$ in this limiting case.

The relationship $\la\rb_1\cdot \rb_{1+k}\ra\sim \omega^{k}$ indicates that the correlations between successive free paths decay exponentially.
\sbyr{These short-range correlations imply that the diffusion of the molecule is normal \cite{Bouchaud1990}: at large times, $\la R^2\ra$ increases linearly with time  and the distribution of the position $R(t)$ is Gaussian.
The fast exponential decay of the correlations means that }
after a few collisions almost no memory of the initial displacement $\rb_1$ remains, and  the preferred direction of movement $\rb_1$ becomes barely   noticeable.  Let us assume our measurement resolution for  $\la\rb_1\cdot \rb_{1+k}\ra$ is $\epsilon$. This implies that after $n=\ln(\epsilon)/\ln \omega$ collisions we are not longer able to detect correlations between successive free paths $\rb_{1+k}$ and the initial displacement $\rb_1$. Within this resolution, $\la R^2(n)\ra=n\ell_e^2$ holds, and therefore $\la R^2(N)\ra/  N =\la R^2(n)\ra/n=\ell_e^2 $. In other words, if we observe the intruder movement  with a spatial resolution of the order $\la R^2(n)\ra^{1/2}$  (or equivalently, with a temporal resolution of around $n\la \tau \ra$), then  what we see is a self-similar random walk.  Obviously,  this self-similarity does not hold if we scrutinize the particle's movement with a finer spatial/temporal resolution. In that case, we perceive  persistence (i.e,  the correlations  between successive steps), something that goes unnoticed when the motion is viewed at larger scales.

\begin{figure}
\begin{center}
\includegraphics[width=.95\columnwidth]{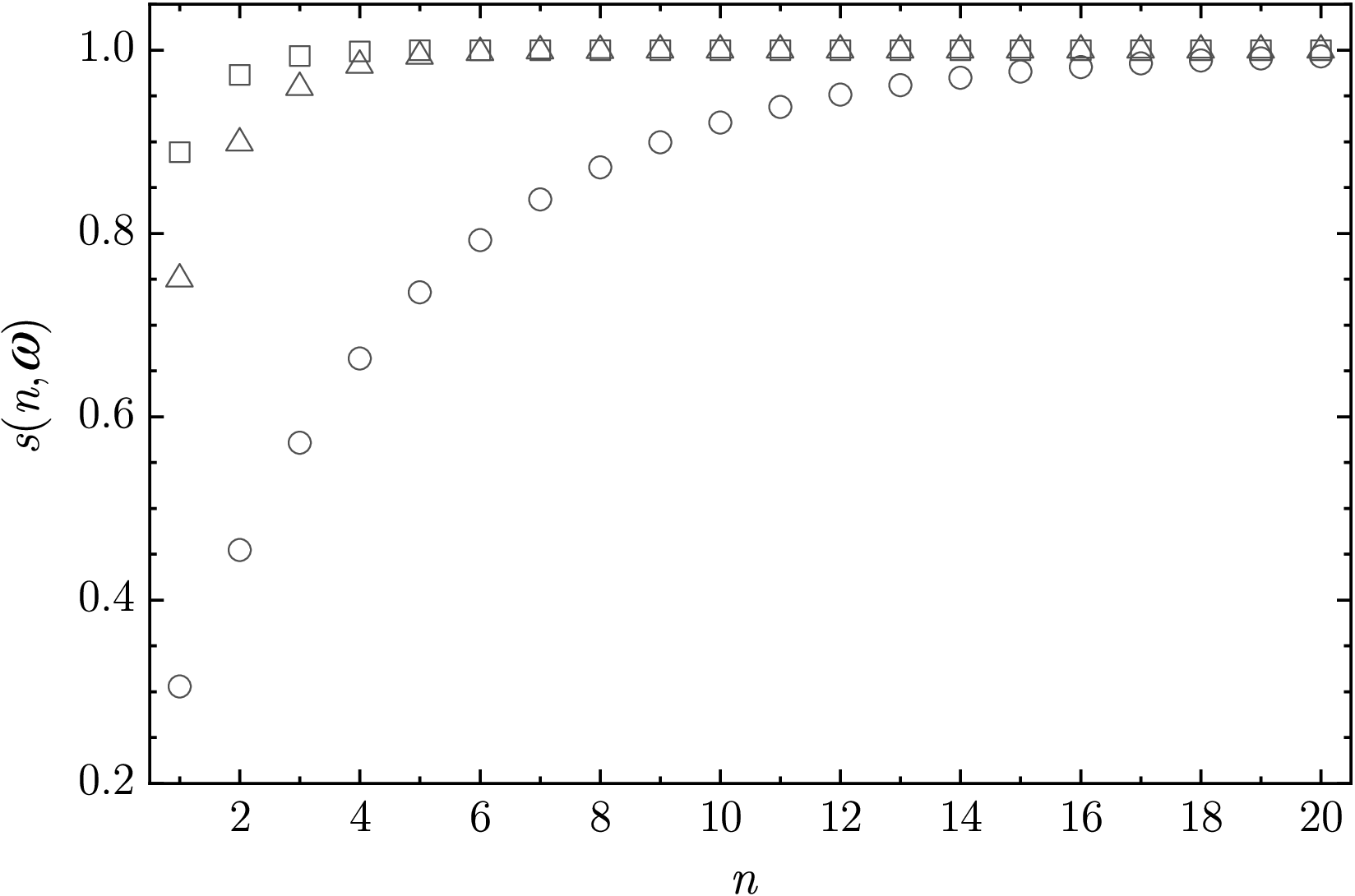}
\end{center}
\caption{\sbyr{Plot of  $s(n,\omega)$ (the slope of the MSD  in units of the asymptotic long-time slope)  vs the number of collisions $n$ for $m_1/m_2=0.5$ ($\omega\simeq 0.245$; square), $m_1/m_2=1$ ($\omega\simeq 0.406$; triangles),  and $m_1/m_2=5$ ($\omega\simeq 0.785$; circles).  } }
\label{fig:Nonlinear}
\end{figure}

\sbyr{On shorter spatial or temporal scales, the ``diffusion'' of the intruder is not normal, i.e. $\la R^2(n)\ra$ is not proportional to the number $n$ of steps (collisions). This conclusion can be easily obtained from Eqs.\ \eqref{MSDles2} and \eqref{sup2}:
\begin{equation}
\label{R2nnl}
\la R^2(n)\ra\approx \la r_1^2\ra \sum_{k=0}^{n-1} (n-k) \omega^k=
\frac{\la r_1^2 \ra }{1-\omega}\,
\left[ n - \frac{\omega(1-\omega^n)}{1-\omega}
\right].
\end{equation}
Equation \eqref{R2nnl} clearly shows that $\la R^2(n)\ra$ is not a linear function of $n$.
However, after a relatively small number $n$ of collisions, the exponential corrective term becomes negligible with respect to $n$, and one eventually finds that the MSD of the molecule grows linearly with $n$. The number $n$ of collisions for which the correction term becomes negligible is smaller when $\omega$ is smaller, i.e., the smaller the mass of the intruder (this is because $\omega$ is an increasing function of the mass of the intruder).
In this normal diffusive regime, the slope $d\la R^2(n)\ra/dn$ of the MSD does not depend on $n$. In our case, the slope of the MSD for all $n$  (in units of the asymptotic long-term slope $\la r_1^2 \ra /(1-\omega)$)  is approximately given by this non-constant function:
\begin{equation}
\label{snome}
1 + \frac{\omega^{1+n}\ln\omega}{1-\omega} \equiv s(n,\omega).
\end{equation}
The separation of  $s(n,\omega)$  from 1 is a measure of how far the intruder's displacement  is from the long-time asymptotic normal diffusion regime. The function $s(n,\omega)$ is plotted
in Fig.~\ref{fig:Nonlinear} versus $n$ for several values of $\omega$.
We see that $s(n,\omega)$ approaches 1 very quickly, meaning that the normal diffusive regime (where $s(n,\omega)=1$) is reached after a small number of collisions. For example, for $m_1/m_2=0.5$, 1, and 5, it takes only about 4, 7, and 28 collisions, respectively, to reach a regime where the slope of $\la R^2(n)\ra$ differs by less than one part in a thousand from its long-term asymptotic value $\la r_1^2 \ra /(1-\omega)$. In other words, if one observes the motion of the intruder on time and/or space scales larger than those corresponding to 4, 7, and 28 collisions, then the motion of the intruder with $m_1/m_2=0.5$, 1, and 5, respectively, will look like normal diffusion.
}
\sbyr{Of course, these numbers depend on the criterion used to decide when normal diffusive behavior is finally achieved. For example, if we use an alternative criterion and decide that the molecule reaches normal diffusion when the value of $d \ln \la R^2(n)\ra/d\ln n$ is no more than, say, 5$\%$ away from 1, then the corresponding numbers of collisions would be 7, 14, and 77, respectively.
}

Finally, since Eq.~\eqref{sup2} plays a key role in our approximations, verifying its validity is a must. To do this, we have to evaluate the average $\langle \mathbf{r}_1 \cdot \mathbf{r}_{1+k} \rangle$. This will be the subject of Sec.\ \ref{sec:5}.

\section{Evaluation of  $\la \rb_1\cdot \rb_{k}\ra$  and diffusion coefficient}
\label{sec:5}

\subsection{Probability of a given path of  $N$ steps}

Let
$
P_N(\vb_1,\tau_{1};\vb_2,\tau_{2};\ldots;\vb_N,\tau_{N})   d\vb_1 d\vb_2 \cdots  d\vb_N$ $d\tau_{N}
$
be the probability of finding molecule 1 (the intruder) with any velocity  which is deflected by collision into $\vb_1$ and $\vb_1 + d\vb_1$ (let's denote it as $\vb_1,d\vb_1$) and travels freely for a time $\tau_{1}$, is then deflected again into $\vb_2,d\vb_2$ and continues to travel freely for another time $\tau_{2}$, and so on until it is deflected  into $\vb_N,d\vb_N$ before finally colliding in the time interval $\tau_{N},d\tau_{N}$. Note that,  because  $\rb_i=\vb_i\tau_{i}$,  this probability  can also be seen as the probability that  a molecule follows the path $\rb_1+\rb_2+\cdots+\rb_N$ with velocities $\{\vb_1,\vb_2,\ldots,\vb_N\}$. The pdf $P_N(\vb_1,\tau_{1};\vb_2,\tau_{2};\ldots;\vb_N,\tau_{N})\equiv P_N(1,2,\ldots;N)$ is given by \cite{Yang1949}
\begin{align}
P_N(1,2,\ldots;N)=&
 f(1)\mu(1)\;  \prod_{i=1}^{N-1}W(i,{i+1}) \nonumber \\
&\times\prod_{j=1}^{N}e^{-\mu(j)  \tau_{j}} \, \mu(N)
\label{PN1N}
\end{align}
where $\mu(i)\equiv \mu(\vb_i)$ is the collision frequency for particles of velocity $\mathbf{v}_i$ and
\begin{equation}
\label{fMaxwell}
    f(1)\equiv f(\vb_1)= \left(\frac{m_1}{2\pi k_\text{B} T}\right)^{3/2} \, e^{-m_1 v_1^2/2k_\text{B}T}
\end{equation}
is the equilibrium pdf of the velocity of the intruder \footnote{
Note that the meaning of $f(\vb)$ here differs from that in Ref~\cite{Yang1949}:  $f(\vb)$ is  what is called $f_1(\vb)/n_1$ in  Ref~\cite{Yang1949}, $n_1$ being the number density of molecule 1}. In addition, $k_\text{B}$ is the Boltzmann's constant and  the transition rate $W(i,{i+1})\equiv W(\vb_i,\vb_{i+1})$ is the probability that a molecule  with precollisional velocity $\vb_1$ is deflected into $\vb_2,d\vb_2$ (postcollisional velocity).  With the help of the equation
\begin{equation}
\label{xxx}
\int f(\vb_1)W(\vb_1,\vb_2)d\vb_1=
f(\vb_2)\mu(\vb_2)
\end{equation}
(which embodies the principle of detailed balancing \cite{Yang1949}), it can be proved that $P_N$ is not normalized to the unity, but to the average collision rate:
\begin{equation}
\int P_N(1,2,\ldots;N) \,   d\vb_1 \cdots \, d\vb_N  d\tau_{N}=  \langle \mu \rangle.
\end{equation}
We will use the notation  $\tilde P_N=P_N/ \langle \mu \rangle$ for the pdf normalized to the unity.

\subsection{\sbyr{The mean square and the pdf of the free path for hard spheres} }
\label{sec:r12}

From the definition of $\tilde P_1$, the mean square free path is defined as
\begin{equation}
\label{r1a}
\langle r_1^2\rangle=   \int_{0}^\infty d \tau_1 \int d\mathbf{v}_1 \,\tilde P_1(\vb_1,\tau_{1}) (v_1\tau_{1})^2.
\end{equation}
When $N=1$, Eq.\ \eqref{PN1N} leads to
\beq
\label{new1}
P_1(\vb_1,\tau_{1})=f(v_1) \mu^2(v_1) e^{-\mu(v_1)\tau_1}.
\eeq
Using the expression \eqref{new1} into Eq.~\eqref{r1a} one easily finds
\begin{equation}
\label{varRL}
\langle r_1^2\rangle= \frac{2}{ \langle \mu \rangle} \int d\vb_1\, f(\vb_1)   \frac{v_1^2}{\mu(v_1)}=\frac{8\pi}{ \langle \mu \rangle} \int_{0}^\infty d v_1\, f(v_1)   \frac{v_1^4}{\mu(v_1)}.
\end{equation}
Since the term $v_1^2/\mu(v_1)$ can be rewritten as $v_1^2\tau_1$ in  Eq.~\eqref{varRL}, then $\langle r_1^2\rangle$ is the average over the velocity distribution $f(v_1)$ of $2  v_1^2\tau_1 / \langle \mu \rangle$:
\begin{equation}
\label{r12MJ}
\langle r_1^2\rangle=2\frac{\la v_1^2\tau_1\ra}{ \langle \mu \rangle}=2\la v_1^2\tau_1\ra  \langle \tau_1 \rangle.
\end{equation}
This justifies our claim in Sec.~\ref{sec:4} that the random walk formula \eqref{DRW} and the   Jeans formula~\eqref{DelleD0pJ} are equivalent.

\begin{figure}
\begin{center}
\includegraphics[width=.95\columnwidth]{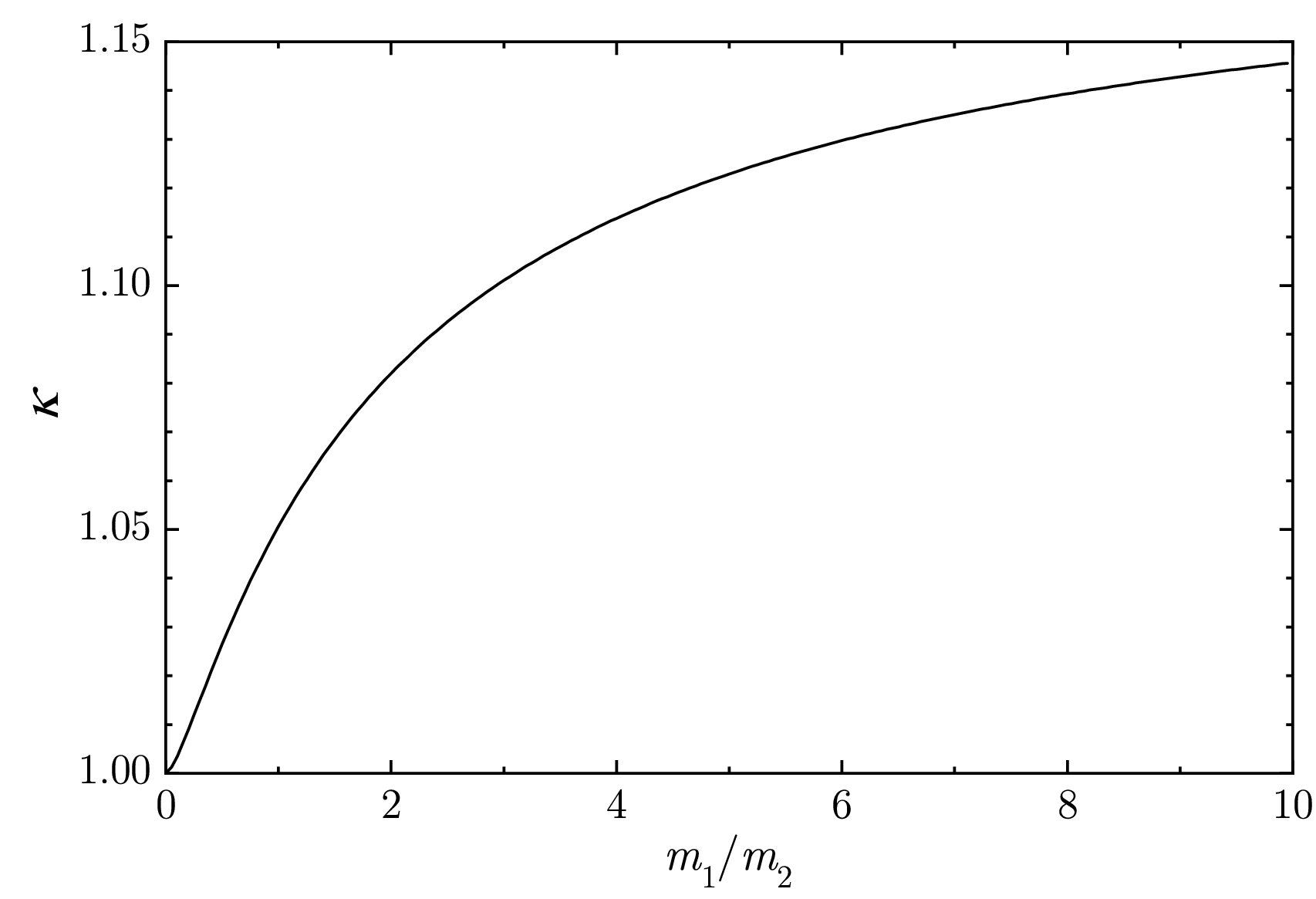}
\end{center}
\caption{Plot of $\kappa={\langle r_1^2\rangle}/{2\lambda^2} $   vs the mass ratio $m_1/m_2$.  }
\label{fig:kappa}
\end{figure}

\sbyr{Introducing the expressions of $f_1$ and $\mu$ for hard spheres given by Eqs.~\eqref{fMaxwell} and \eqref{mueHS} into Eq.~\eqref{varRL}, one finds that the mean square free path $\langle r_1^2\rangle$}  (in units of $2\lambda^2$)   is
\begin{equation}
\label{kappa}
\kappa\equiv\frac{\langle r_1^2\rangle}{2\lambda^2} =2\sqrt{\pi}\, \left(\frac{m_1}{m_2}\right)^2 \left(1+\frac{m_1}{m_2}\right)^{1/2}\; I_{1,1}(m_1/m_2)
\end{equation}
where
\begin{equation}
\label{I11}
I_{1,1}(m)=\int_0^\infty e^{-m y^2}   \frac{y^4}{E(y)} dy.
\end{equation}
No explicit expression for $I_{1,1}(m)$ is known, but it can be easily evaluated numerically. The function $\kappa$ is plotted in Fig.~\ref{fig:kappa} as a function of the mass ratio $m_1/m_2$.
We see that the larger $m_1/m_2$, the larger is the difference of $\kappa$  from 1. The largest difference  is for  $m_1/m_2\to \infty$. In this case $\kappa=3\pi/8\simeq 1.1781$ \footnote{To evaluate $\kappa$ for $m_1/m_2\to\infty$ one first notes that, due to the term $\exp(-m y^2)$ in the integrand of $I_{1,1}(m)$, the only relevant values of $y$ for the evaluation of  this integral are those around zero.
But $\lim_{y\to 0}E(y)=2$ so that the dominant term of $I_{1,1}$ is given by half the integral of $y^4 \exp(-m y^2)$. This integral is equal to $3\sqrt{\pi}/8m^{5/2}$, and the value of $\kappa=3\pi/8$, for $m_1/m_2\to\infty$, follows.
}.

\sbyr{Note} that
\begin{equation}
\label{P1taur1}
P_1(\vb_1,\tau_{1})d\vb_1d\tau_{1}=(1/v_1)P_1(\vb_1,r_1/v_1)d\vb_1 dr_1.
\end{equation}
Using the standard notation $\ell$ for the length $r_1$ of the free path, and taking into account Eqs.~\eqref{new1} and \eqref{P1taur1},  we find that the pdf of a free path of length $\ell$ \sbyr{(whatever its corresponding velocity)} is given by
\begin{equation}
\label{Pell}
P(\ell)=\int \frac{\tilde{P}_1(\vb,\ell/v)}{v} d\vb =
 \frac{1}{\la \mu\ra}\int  f(\vb) \frac{ \mu^2(v)}{v}   e^{-\mu(v)\ell/v}  d\vb.
\end{equation}
From this expression, we can evaluate $\la \ell^2\ra \equiv \langle r_1^2\rangle=\int \ell^2 P(\ell) d\ell$. Changing the order of integration, the integral over $\ell$ is immediate and we reobtain Eq.~\eqref{varRL}.
A different way of obtaining Eq.~\eqref{Pell} was used in Ref.~\cite{Visco2008}.  This way involves the so-called on-collision pdf velocity distribution $f_\text{coll}=f(v) \mu(v)/ \la \mu\ra$ (see Refs.~\cite{Lue2005,Visco2008,Paik2014}). An interesting short discussion of $P(\ell)$ can be found in Section II.C of Ref.~\cite{Visco2008} (see also Ref.\ \cite{Paik2014}).

\sbyr{It must be remarked that the quantity} $\kappa$ would be 1 for all $m_1/m_2$ if the simple approximation $P(\ell)\approx \exp(-\ell/\lambda)/\lambda$ for the  free path length distribution Eq.~\eqref{Pell} would hold.  This was the approximation made to obtain $D_\text{eKT}$ (see the end of Sec.~\ref{sec:2}) as well as   the Jeans formula \eqref{DJ} for the diffusion coefficient $D_\text{J}$ (see the beginning of Sec.~\ref{sec:3}). Of course, this is only an approximation.
The exact value of $P(\ell)$ for hard spheres can be obtained from Eq.~\eqref{Pell} as
\begin{equation}
    \label{xd}
   \lambda\, P(\ell)  = \frac{2(m_1/m_2)^2}{\pi (1+m_1/m_2)}\, I_{\ell}(\ell/\lambda,m_1/m_2).
   \end{equation}
\sbyr{Here, the function $I_{\ell}(x,m)$ is given by}
\begin{equation}
\label{xe}
 I_{\ell}(x,m)=\int_0^\infty dy\, y\, E^2(y) \,
     \exp\left[-m y^2-\frac{E(y)}{y\sqrt{\pi (1+m)}}\,x\right],
\end{equation}
\sbyr{where the quantity $E(y)$ is defined by Eq.\ \eqref{Edef}.}

\begin{figure}
\begin{center}
\includegraphics[width=.95\columnwidth]{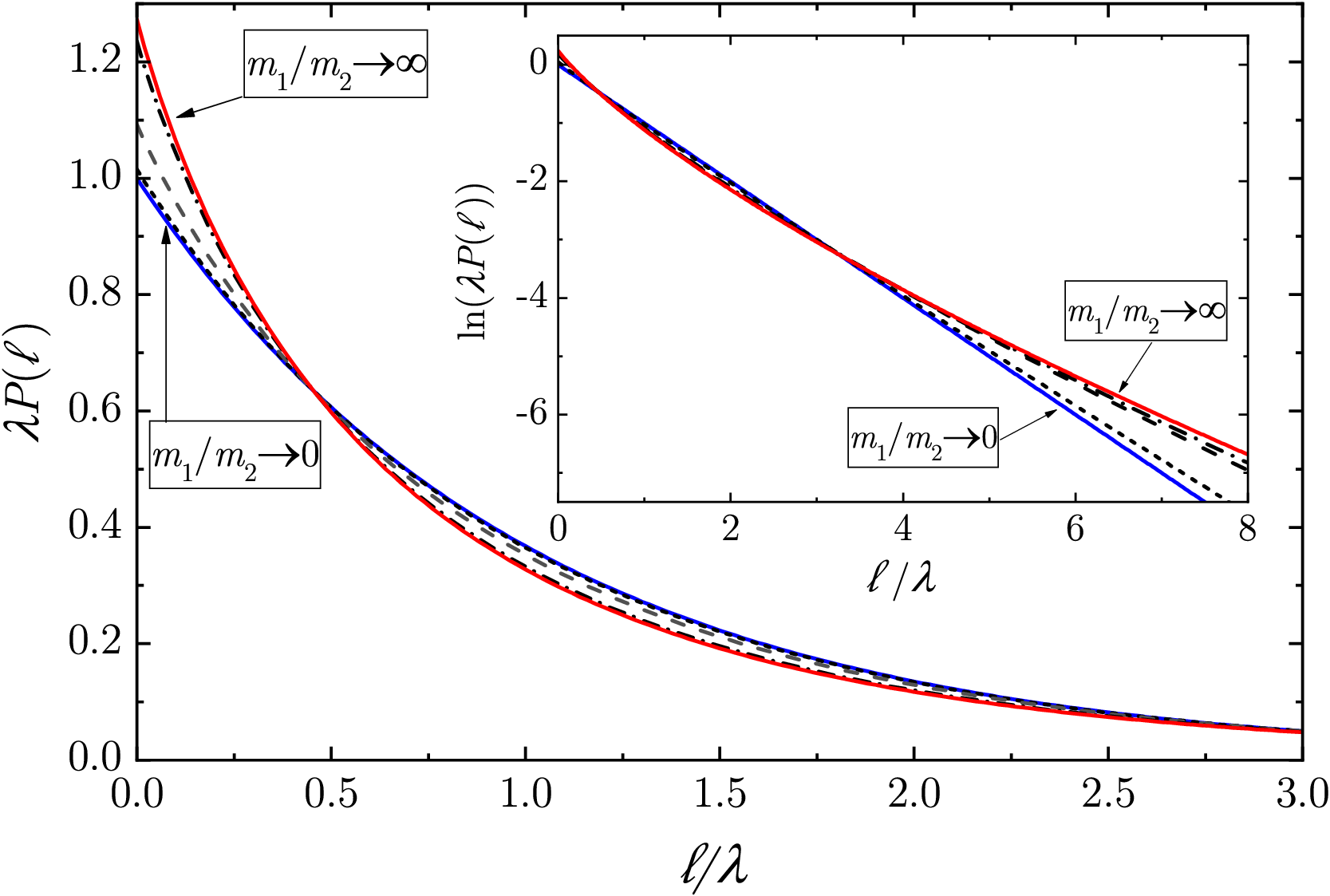}
\end{center}
\caption{Plot of  $\lambda P(\ell)$ for hard spheres  for   $m_1/m_2=0.2$ (dotted line), $m_1/m_2=1$ (dashed line), $m_1/m_2=10$ (dashed-dotted line). The approximation valid for a very light intruder, $\exp(-\ell/\lambda)/\lambda$,
\sbyr{and for a very massive intruder, Eq.~\eqref{PellmLarge}, are the solid lines}.
Same in the inset but as a semi-log plot.}
\label{fig:Pell}
\end{figure}

The (dimensionless) function $\lambda P(\ell)$ is plotted in Fig.~\ref{fig:Pell} as a function of $\ell/\lambda$ for several values of the mass ratio $m_1/m_2$. It is quite apparent that Fig.~\ref{fig:Pell} confirms that the expression $\exp(-\ell/\lambda)/\lambda$ is only an approximation of the true length distribution  $P(\ell)$, as expected. Figure \ref{fig:Pell} also highlights that the approximation $P(\ell)\approx\exp(-\ell/\lambda)/\lambda$ becomes more accurate as the mass ratio $m_1/m_2$ decreases.
\sbyr{From a physical point of view}, this is an expected trend since $P(\ell)\approx\exp(-\ell/\lambda)/\lambda$ characterizes the free path distribution of an intruder moving in a gas composed of very massive molecules, which are, from the intruder's perspective, almost at rest.
\sbyr{Mathematically, this result is obtained from the asymptotic expansion of $P(\ell)$ for $m\equiv m_1/m_2\to 0$ (see Appendix \ref{ap:Asymptotic}).  On the other hand, the asymptotic expansion of $P(\ell)$ for $m\to \infty$ (very heavy intruder) leads to (see Appendix \ref{ap:Asymptotic})
\begin{equation}
\label{PellmLarge}
\lambda P(\ell)
\sim
\frac{4}{\pi^{3/2}} \displaystyle G_{0,3}^{3,0}\!\left(\left.\begin{array}{c}
\,\\ 0,\frac{1}{2},1
\end{array}\;\right|\,\frac{\ell^2}{\pi\lambda^2}\right),
\end{equation}
where $G_{0,3}^{3,0}$ is a Meijer G-function \cite{WolframMeijerG}. We have also plotted this expression in Fig.~\ref{fig:Pell}. It is remarkable that the line corresponding to the expression \eqref{PellmLarge} is almost indistinguishable from the one obtained from Eq.\ \eqref{xd} for $m=10$.
}

\subsection{The average $\langle \rb_1\cdot \rb_{k}\rangle$ and the diffusion coefficient for hard spheres}
\label{sec:aveDHS}


We aim to evaluate the average $\la r_1 r_k \cos\theta_{1,k}\ra$.  Following the same type of steps as those carried in Sec.~\ref{sec:3} [using the spherical law of cosines, Eq.~\eqref{cos1k}], one finds the result
\begin{equation}
\label{xx}
\la \rb_1\cdot\rb_k \ra= \la r_1 r_k \cos\theta_{1,2}\cos\theta_{2,3}\ldots\cos\theta_{k-1,k}\ra.
\end{equation}
More explicitly,
\begin{align}
\langle \rb_1\cdot\rb_k \rangle=&   \int \tilde P_k(1,\ldots,k)\,  v_1 v_k\,\tau_{1} \tau_{k}
 \cos\theta_{1,2}\ldots\cos\theta_{k-1,k}\nonumber\\
 &\times d\vb_1 d\tau_{1}\ldots d\vb_k d\tau_{k}.
\label{r1rka}
\end{align}
Using Eq.~\eqref{PN1N} and integrating over all $\tau_i$, Eq.~\eqref{r1rka} becomes
\begin{align}
\langle \rb_1\cdot\rb_k\rangle=&\frac{1}{\langle \mu \rangle}
\int f(\vb_1)   \frac{W(\vb_1,\vb_2)\cdots W(\vb_{k-1},\vb_k)}{\mu(v_1)\cdots\mu(v_k)} \nonumber\\
  &\times \, v_1 v_k\,  \cos\theta_{1,2}\cdots\cos\theta_{k-1,k}\, d\vb_1 \cdots d\vb_k.
\label{r1rk}
\end{align}


Using the collision rate \eqref{mueHS} and the transition rate \eqref{WeHS} of  hard spheres in Eq.~\eqref{r1rk} and working in spherical coordinates, one finds that $\langle \rb_1\cdot \rb_{k}\rangle$ can be written as
\beqa
\label{r1rkesf}
\langle \rb_1\cdot\rb_k\rangle
&=&\frac{2^{2-k}}{\pi^{3/2}} \left(\frac{m_1}{m_2}\right)^{2}\, \left(1+\frac{m_1}{m_2}\right)^{2k-5/2}\, \frac{1}{(n_{2}\sigma_{12}^2)^2}
\nonumber\\
& & \times I_{1,k}(m_1/m_2),
\eeqa
for $k\ge 2$, where
\beqa
\label{I1k}
I_{1,k}(m)
&=&
\int_0^\infty\cdots\int_0^\pi  e^{-m y_1^2}\,
\frac{\widetilde{W}(\yb_1,\yb_2)\cdots \widetilde{W}(\yb_{k-1},\yb_k)}{E(y_1)\cdots E(y_k)}\nonumber\\
&\times&   y_1^3y_2^2 \cdots y_{k-1}^2 y_k^3  \cos\theta_{12}\cdots \cos\theta_{k-1,k} \nonumber\\
&\times& \sin\theta_{12}\cdots \sin\theta_{k-1,k}  dy_1 \cdots dy_k d\theta_{12}\cdots d\theta_{k-1,k}.
\nonumber\\
\eeqa
In units of $\lambda^2$, one gets
\beqa
\label{r1rkla2B}
\frac{\langle \rb_1\cdot\rb_k\rangle}{\lambda^2}
&=&2^{2-k}\sqrt{\pi} \left(\frac{m_1}{m_2}\right)^{2}\,  \; \left(1+\frac{m_1}{m_2}\right)^{2k-3/2}\nonumber\\
& & \times
I_{1,k}(m_1/m_2),
\eeqa
for $k\ge 2$. Therefore,  according to Eq.\ \eqref{ell2redSer}, we find that the collisional series can be written as
\beq
\label{xg}
\ell_e^2=\sum_{k=1}^\infty \frac{2^{3-k}}{\pi^{3/2}} \left(\frac{m_1}{m_2}\right)^{2}\, \left(1+\frac{m_1}{m_2}\right)^{2k-5/2}\, \frac{I_{1,k}(m_1/m_2)}{(n_{2}\sigma_{12}^2)^2}
\eeq
Alternatively, taking into account Eqs.~\eqref{kappa} and \eqref{r1rkla2B}, the $k$-th term of the reduced collisional series $c_k=2\la  \rb_1\cdot \rb_k \ra/\la r_1^2 \ra$  is given by
\beqa
\label{ckelasHS}
c_k&=&\frac{\la  \rb_1\cdot \rb_k \ra}{\kappa \lambda^2}
\nonumber\\
&=&
2^{1-k}\left(1+\frac{m_1}{m_2}\right)^{2(k-1)}\; \frac{I_{1,k}(m_1/m_2)}{I_{1,1}(m_1/m_2)}.
\eeqa
Note that the expression \eqref{ckelasHS} also applies for $k=1$. Hence,   the reduced collisional series \eqref{ell2redSer}  can be written as
\begin{align}
\label{xh}
\frac{\ell_e^2}{\langle r_1^2\rangle}
&= \sum_{k=1}^\infty  2^{1-k}\left(1+\frac{m_1}{m_2}\right)^{2k-2}\; \frac{I_{1,k}(m_1/m_2)}{I_{1,1}(m_1/m_2)}.
\end{align}
From Eq.~\eqref{DelleD0Ser},  the diffusion coefficient $D$ is
\begin{equation}
\label{xseries}
D=D_{\text{eKT}}\;\frac{ \langle r_1^2\rangle }{2\lambda^2}\, \sum_{k=1}^\infty  2^{2-k}\left(1+\frac{m_1}{m_2}\right)^{2k-2}\; \frac{I_{1,k}(m_1/m_2)}{I_{1,1}(m_1/m_2)}.
\end{equation}
Of course, the problem with the series \eqref{xseries} is that the $k$-th term requires the evaluation of  $I_{1,k}$, namely, a $2k-1$-dimensional integral which is  unapproachable   for large $k$.  However, the numerical evaluation of  the first few  terms is  accessible. We show this in the following Section~\ref{sec:compa}.

\sbyr{Before finishing this section it is worthwhile to compare the integral forms of $\langle \rb_1\cdot\rb_k\rangle$ and the mean persistence ratio $\omega$.  We start noticing that}
the integrand of Eq.~\eqref{r1rk} has an interesting form. Let us denote this integrand by the symbol $A_{1,k}$, that is,
\begin{equation}
\label{xa}
\langle \rb_1\cdot\rb_k\rangle=\int A_{1,k} d\vb_1 \cdots d\vb_k.
\end{equation}
We see that the integrand of $\langle \rb_1\cdot\rb_k\rangle$ and the integrand of  $\langle \rb_1\cdot\rb_{k-1}\rangle$ just differ by the factor $p_{k-1,k}$, that is $A_{1,k}=A_{1,k-1}p_{k-1,k}$, where
\begin{equation}
\label{xb}
p_{k-1,k}= \frac{W(k-1,k)}{\mu(k)} \frac{v_{k}}{v_{k-1}} \cos\theta_{k-1,k}.
\end{equation}
The form of $A_{1,k}=A_{1,k-1}p_{k-1,k}$    suggests that $\la \rb_1\cdot\rb_{k-1}\ra$,  $\la \rb_1\cdot\rb_{k}\ra$, etc.,  might follow a sequence with a  ratio between consecutive terms related to the integration of $p_{k-1,k}$.  Note the  closeness of $p_{k-1,k}$ to the integrand of the integral
\begin{equation}
\label{xc}
 \omega = \int f(v_{k-1})  \frac{W(v_{k-1},v_{k})}{ \langle\mu \rangle}  \frac{v_{k}}{v_{k-1}}\cos\theta_{k-1,k}  d\vb_{k-1} d\vb_k
\end{equation}
that provides the mean persistence ratio.  This observation suggests the proportionality relationship between the ratio of two consecutive terms of the collisional series and the  mean persistence ratio that we discussed in Secs.~\ref{sec:3} and \ref{sec:4}.

\section{Comparison with DSMC simulations and Boltzmann results for hard spheres}
\label{sec:compa}

In this Section we estimate the terms $c_k$ of the reduced collisional series by numerically integrating $I_{1,k}$; this result will be assessed via a comparison with DSMC simulations. Additionally, the theoretical predictions for the diffusion coefficient given by Eqs.~\eqref{DJ}  and  \eqref{DRW} and by the first- and second-Sonine approximations (one and two terms in the Sonine polynomial expansion of the velocity distribution function) to the Chapman--Enskog solution \cite{CC70} to the Boltzmann equation will be also compared with simulation data obtained from the DSMC method \cite{B94}.
 It is important to remark that both DSMC simulations and Sonine approximations are derived through the Boltzmann kinetic equation, which plays a pivotal role in modeling the kinetic behavior of gaseous particles. This approach provides a solid theoretical foundation for our comparisons, enabling us to assess the validity and accuracy of our predictions in relation to well-established and widely accepted methods for describing transport phenomena in gaseous systems.

\begin{figure}
\begin{center}
\includegraphics[width=.95\columnwidth]{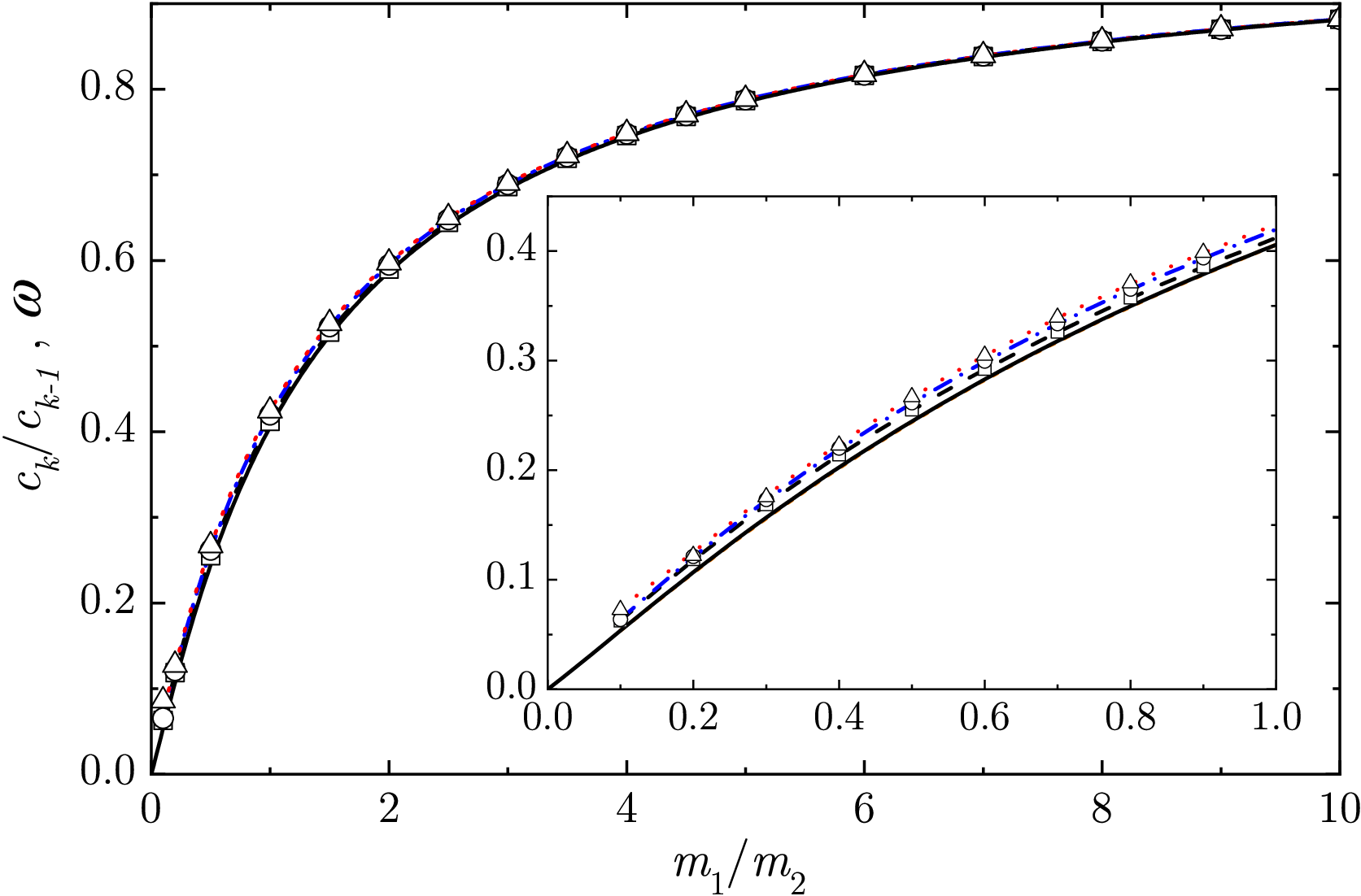}
\end{center}
\caption{Plot of  $c_2/c_1$ (dashed line, squares), $c_3/c_2$  (dashed-dotted line,circles) and $c_4/c_3$ (dotted line, triangles)  vs the mass ratio $m_1/m_2$.  The lines are obtained by numerically evaluating the expression \eqref{ckelasHS} for each mass ratio  while the symbols are  DSMC simulation results. The solid line is the mean persistence ratio $\omega$. Inset: the same  when the intruder is lighter than the molecules of the gas.
\label{fig:ck}}
\end{figure}

In Table \ref{tabla1}, we present  numerical estimates for the first four terms of the collisional series \eqref{ell2Sume} for various representative values of the mass ratio $m_1/m_2$.  These estimates require the numerical integration of  $I_{1,k}$ [see Eqs.~ \eqref{kappa}, \eqref{I11}, \eqref{r1rkesf}, and \eqref{I1k}]. As $k$ increases, the numerical integration becomes progressively challenging, leading to less precise results, particularly for small values of $m_1/m_2$. We also include in Table \ref{tabla1} values obtained through DSMC simulations at very low densities. Some technical details on the implementation of the DSMC method to the evaluation of the averages $\la  \rb_1\cdot \rb_k \ra$ are provided in the Appendix \ref{ap:dsmc}. \sbyr{Specifically, we consider approximately $10^5$ molecules, conducting up to $5 \times 10^5$ runs, each comprising an average of $500$ proposed candidates to collide in every Monte Carlo step, with the final number of collisions depending on the mass ratio $m_1/m_2$.} Table \ref{tabla1} highlights the excellent agreement between DSMC simulations and the numerical results.  From these results, the values of the first ratios between successive terms of the reduced collisional series can be derived. These values are given in Table \ref{tabla2}. Additionally, we provide the corresponding values of the mean persistence ratio $\omega$. The proximity of the coefficients $c_k/c_{k-1}$ to $\omega$ is quite remarkable.
Note that this closeness improves for heavier intruders.
In Fig.~\ref{fig:ck} we compare the first three ratios of the reduced collisional series, $c_2/c_1$, $c_3/c_2$ and $c_4/c_3$,  with $\omega$ for a large set of mass ratios.  The agreement  is stunning, providing strong support for the approximation $c_{1+k}/c_k\approx \omega$ or, equivalently, for $c_{1+k} \approx \omega^k$ (recall that $c_1=1)$. This result is the cornerstone of the approximations  \eqref{R2ome} and \eqref{R2omep} for the MSD,  and of the approximations \eqref{DJ} and \eqref{DRW} for the diffusion coefficient. These findings highlight
the reliability of the random walk approach.

\begin{table}
\begin{tabular}{|c|c|c|c|c|}
 \hline
  $ m_1/m_2$ & $\langle r_1^2 \rangle$ & $\langle \rb_1\cdot \rb_2\rangle$ & $\langle \rb_1\cdot \rb_3\rangle$ & $\langle \rb_1\cdot \rb_4 \rangle$ \\
  \hline
0.2 & 2.0179   & 0.1198  & 0.0145   & 0.0018  \\
    & 2.0174   & 0.1198    & 0.0145  & 0.0018 \\
\hline
0.5 & 2.0529    & 0.2621     & 0.0686  & 0.0183  \\
    & 2.0524    & 0.2619    & 0.0686  & 0.0183 \\
\hline
1   & 2.1012   & 0.4331   & 0.1818 &  0.0772   \\
    & 2.1009    & 0.4329    & 0.1818  & 0.0772 \\
\hline
2   & 2.1640    &  0.6386   &  0.3799 &  0.2269   \\
    & 2.1636    &  0.6383   &  0.3798 &  0.2268  \\
\hline
5   & 2.2458   &  0.8838   &  0.6963  &  0.5489   \\
    & 2.2454    &  0.8833   &  0.6960 &   0.5487   \\
\hline
10  & 2.2915   &  1.0103  &  0.8910 &  0.7858   \\
    & 2.2911    &  1.0100   &  0.8907 &  0.7856   \\
 \hline
\end{tabular}
\caption{
Numerical and  DSMC  simulation values of  $\langle \rb_1\cdot \rb_k \rangle$, with $k=1,2,3,4$,  in units of $\lambda^2$ for several values of $m_1/m_2$. For each value of the mass ratio $m_1/m_2$, the first row are the numerical values and the second row are the   DSMC  values.}
\label{tabla1}
\end{table}

\begin{table}
\begin{tabular}{|c|c|c|c|c|}
 \hline
  $ m_1/m_2$ & $c_2/c_1$  &  $c_3/c_2$  &  $c_4/c_3$   & $\omega$ \\
  \hline
0.2 &  0.119   & 0.121  & \emph{0.126}  & 0.107\\
\hline
0.5 & 0.255  & 0.262   & 0.267  & 0.245 \\
\hline
1   & 0.412   & 0.420   & \emph{0.425} &  0.406\\
\hline
2   & 0.590  &  0.595 &  0.597 &  0.587\\
\hline
5   & 0.787   &  0.788 &  0.788  &  0.785 \\
\hline
10  & 0.882   &  0.882  &  0.882  &  0.881 \\
 \hline
\end{tabular}
\caption{
The first three ratios $c_2/c_1$, $c_3/c_2$ and $c_4/c_3$  of the  coefficients of the reduced  collisional series   \eqref{ell2redSer}   for several values of $m_1/m_2$ evaluated by means of Eq.~\eqref{ckelasHS}.  DSMC simulations provide the same results for all cases with the exception of $c_4/c_3$ for $m_1/m_2=0.2$ and  $m_1/m_2=1$; in these cases the simulation results are 0.127 and 0.424, respectively. The last column is the value of the mean persistence ratio $\omega$ for hard spheres [see Eq.~\eqref{omega}].
}
\label{tabla2}
\end{table}

It has long been recognized that a connection exists  between the random walk formulation of the diffusion in gases and the Jeans formulation, which involves the persistence of velocity. It is therefore surprising that the possibility (as demonstrated in this paper) of deriving Jeans' diffusion coefficients from a random walk description of molecular diffusion has gone unnoticed.
We suspect  that this oversight may be attributed to the apparent disparity between the two formulations (Jeans and random walk approaches) when relying on Yang's (very rough and poor) estimate of $\langle \rb_i\cdot\rb_{i+k}\rangle$ for hard spheres with $k=1,2$ \cite{Yang1949}.
For $m_1=m_2$, the values of $c_2/c_1$ and $c_3/c_2$ that one obtains from Yang's results (see  Appendix C of Ref.~\cite{Yang1949}) would be $0.72$ and $0.47$, respectively. These values are very different from each other (and from the true values $0.412$ and $0.420$ we give in Table \ref{tabla2}) and from the mean persistence ratio $0.406$ corresponding to this case.
These imprecise estimates by Yang are understandable given the complexity of the integrals $I_{1,k}$, and that Yang lacked the computational means to numerically evaluate them. Had Yang been able to compute the first few terms   $\langle \rb_1\cdot\rb_{1+k}\rangle$ accurately, it is likely that he or others (Monchick, for example: see Refs.~\cite{Monchick1962} and \cite{Monchick1967}, specifically compare section IV of \cite{Monchick1967} with the last paragraph of section IV of \cite{Monchick1962}) would have recognized the close resemblance of the ratio between these terms to the mean persistence ratio. Such a realization would have unveiled the connection between Yang's random walk formulas and those of Jeans.

From the relation $c_2\approx c_{1+k}/c_k\approx \omega$, we have obtained $c_{1+k} \approx \omega^k$. Employing this result in Eqs.\eqref{ell2redSer} and \eqref{DelleD0Ser}, we achieve the expressions \eqref{ele2p} and \eqref{DRW}  in Sec.\ \ref{sec:4}. Note, however, that the approximation $c_2\approx c_{1+k}/c_k$ also implies $c_{1+k} \approx c_2^k$. The use of this result in Eqs.~\eqref{ell2redSer} and \eqref{DelleD0Ser} yields the expressions
\begin{equation}
\label{R2c2p}
\langle R^2(t) \rangle = \langle N(t) \rangle \, \frac{\langle r_1^2 \rangle}{1-c_2},
\end{equation}
and
\begin{equation}
\label{DRWprime}
D = D_{\text{eKT}} \frac{\kappa}{1-c_2} = \frac{\lambda \bar{v}}{d} \frac{\kappa}{1-c_2}
\equiv D'_{\text{RW}}.
\end{equation}
Certainly, choosing for $\omega$ as the estimate for $c_{1+k}/c_k$ offers the advantage of having  the closed explicit expression \eqref{omega}, while  the evaluation of  $c_2$ requires numerical computation. On the other hand, as we will see next, $D'_\text{RW}$ turns out to be somewhat better than $D_\text{RW}$.

In  Table~\ref{tabla3} we compare the diffusion coefficients provided by the random walk approach, Eqs.~\eqref{DJ}, \eqref{DRW} and \eqref{DRWprime}, with the diffusion coefficients provided by two standard approximations of the kinetic theory of gases, namely,  the first and second Sonine approximations. In terms of $D_\text{eKT}$, the diffusion coefficient $D_\text{KT}^{[1]}$ for hard spheres obtained from the first Sonine approximation is given by (see Sec.~14.2 of Ref.\ \cite{CC70})
\begin{equation}
\label{xi}
\frac{D_\text{KT}^{[1]}}{D_\text{eKT}}=  \frac{9\pi}{32} \, \left(1+\frac{m_1}{m_2}\right).
\end{equation}
The second-Sonine approximation $D_\text{KT}^{[2]}$ is  (see Sec.~14.3 of Ref.\ \cite{CC70})
\begin{equation}
\label{xj}
\frac{D_\text{KT}^{[2]}}{D_\text{eKT}}
= \frac{9\pi}{32} \, \left(1+\frac{m_1}{m_2}\right)\, \frac{13+16m_1/m_2+30(m_1/m_2)^2}{12+16m_1/m_2+30(m_1/m_2)^2}.
\end{equation}

\begin{table}  
\begin{tabular}{|c|c|c|c|c|c|c|}
 \hline
  $ m_1/m_2$ & $D_\text{J}$  & $D_\text{RW}$ & $D'_\text{RW}$ & $D_\text{KT}^{[1]}$ &$D_\text{KT}^{[2]}$ &DSMC\\
  \hline
0.2 &  1.12   & 1.13  & 1.14  & 1.06   & 1.12  & 1.15\\
\hline
0.5 & 1.32    & 1.36  & 1.38  & 1.33   & 1.37  & 1.40\\
\hline
1   & 1.68    & 1.77  & 1.79  & 1.77   & 1.80  & 1.82\\
\hline
2   & 2.42    & 2.62  & 2.64  & 2.65   & 2.67  & 2.67\\
\hline
5   & 4.66    & 5.23  & 5.27  & 5.30   & 5.31  & 5.34\\
\hline
10  & 8.40    & 9.63  & 9.70  & 9.72   & 9.72  & 9.84\\
 \hline
\end{tabular}
\caption{
The diffusion coefficients $D_\text{J}$, $D_\text{RW}$, $D'_\text{RW}$, $D_\text{KT}^{[1]}$ and  $D_\text{KT}^{[2]}$, as well as DSMC  simulations values in units of  $D_{KT}^{[0]}$   for several values of $m_1/m_2$.
}
\label{tabla3}
\end{table}

\begin{figure}
\begin{center}
\includegraphics[width=.95\columnwidth]{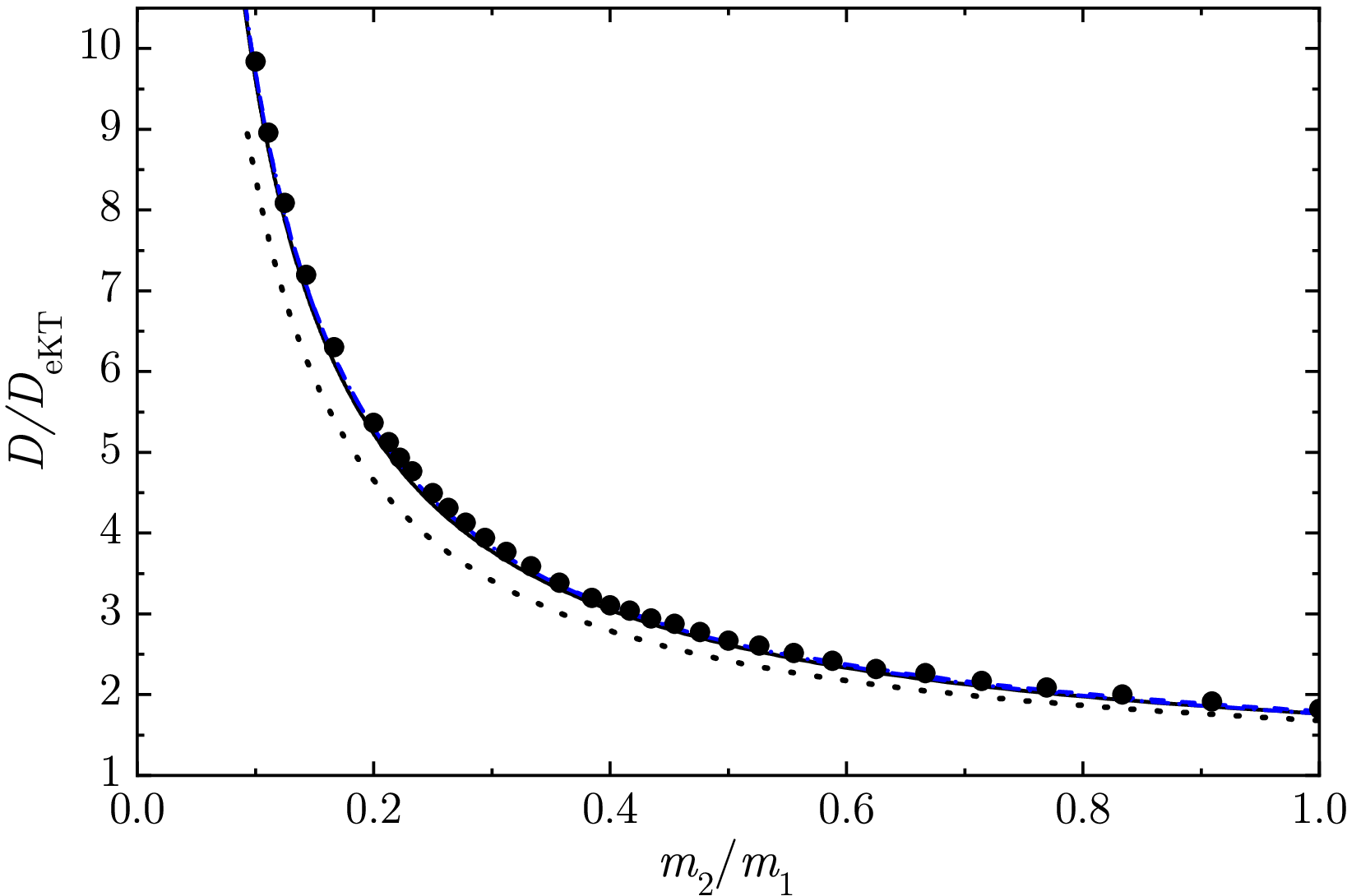}
\end{center}
\caption{Plot of the reduced diffusion coefficient $D/D_\text{eKT}$ as a function of the mass ratio $m_2/m_1$ when the intruder is heavier than the molecules of the gas. The lines correspond to $D=D_\text{J}$ (dotted line), $D=D_\text{RW}$ (solid line), $D=D_\text{KT}^{[1]}$ (dashed-dotted line) and  $D=D_\text{KT}^{[2]}$ (dashed line).  The circles are  DSMC  simulation values.
\label{fig:Dm1large}}
\end{figure}

\begin{figure}
\begin{center}
\includegraphics[width=.95\columnwidth]{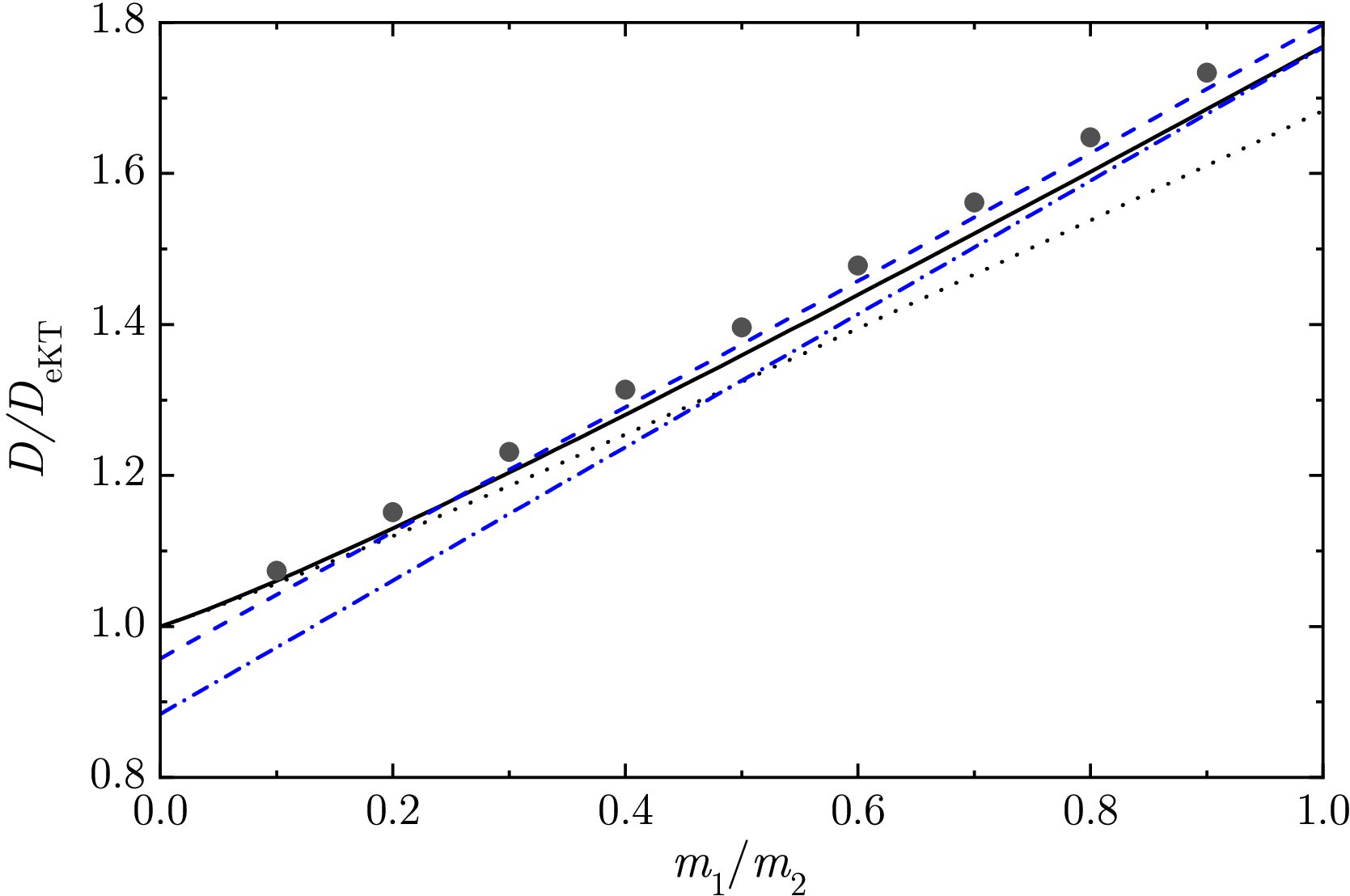}
\end{center}
\caption{Plot of the reduced diffusion coefficient $D/D_\text{eKT}$ as a function of the mass ratio  $m_1/m_2$ when the intruder is lighter than the molecules of the gas. The lines correspond to $D=D_\text{J}$  (dotted line), $D=D_\text{RW}$ (solid line), $D=D_\text{KT}^{[1]}$ (dashed-dotted line) and  $D=D_\text{KT}^{[2]}$ (dashed line).  The circles are  DSMC  simulation values.
\label{fig:Dm1small}}
\end{figure}

We plot in Fig.~\ref{fig:Dm1large} the diffusion coefficients $D_\text{J}$, $D_\text{RW}$, $D_\text{KT}^{[1]}$ and $D_\text{KT}^{[2]}$  as a function of the mass ratio $m_2/m_1$ when the intruder is heavier than the molecules of the gas ($m_1>m_2$). The complementary case of an intruder lighter than the molecules of the gas ($m_1<m_2$)  is considered in Fig.~\ref{fig:Dm1small}.
In both figures \ref{fig:Dm1large} and \ref{fig:Dm1small} we show the results of  DSMC simulations. We see that the simplest formula with persistence, $D_\text{J}$ [Eq.~\eqref{DJ}] is a significant improvement over the diffusion coefficient of the elementary kinetic theory $D_\text{eKT}$ and qualitatively captures the effect of the mass ratio of the random walk coefficient. However, at a quantitative level, this approach is not fully satisfactory. On the contrary, the improved random walk formula [given by the coefficient $D_\text{RW}$, see Eq.~\eqref{DRW}]  provides excellent results. In fact, when the intruder is lighter than the molecule of the gas, it is even better than the first Sonine approximation.  In the opposite case, when the intruder is heavier than the molecule of the gas, the results are quite good, slightly worse than, but comparable, to the results obtained with the Sonine approximations.

\section{Conclusions}
\label{sec:Conclu}

In this paper we have used the fact that mean square displacement of a molecule in a gas  at time  $t$ can be written as the mean square displacement of a molecule with isotropic collisions, $\la R^2(t)\ra=\la  N(t)\ra \la r_1^2\ra$, corrected by the prefactor (the reduced collisional series) $\sum_{k=1}^\infty c_k=1+2\sum_{k=2}^\infty \la \rb_1\cdot\rb_k\ra/\la r_1^2\ra$. The key result of our paper is the realization that $c_{1+k}$  can be well approximated by $\omega^k$, leading to the result
\beq
\label{7.1}
\sum_{k=1}^\infty c_k=\sum_{k=0}^\infty \omega^k=
\frac{1}{1-\omega},
\eeq
\sbyr{and hence,
\beq
\la R^2(t)\ra=\la  N(t)\ra \frac{\la r_1^2\ra}{1-\omega}.
\eeq}
This implies that, to estimate the  MSD of a molecule in a gas at equilibrium, only two quantities need to be known: the mean square free path $\la r_1^2\ra$, and  the  mean persistence ratio $\omega$.   Expressed as $\langle R^2(t) \rangle / (2dt)$, the diffusion coefficient is the value associated with purely isotropic collisions (i.e., $\lambda \bar{v}/d$), corrected by the factor $(1-\omega)^{-1}$.     We have justified this result in two ways: first, through simple and intuitive arguments (that involve some uncontrolled approximations); and second, by rigorously evaluating the first few terms $c_k$ of the reduced collisional series and realizing that, for hard spheres, these terms form an almost perfect geometric sequence with a common ratio very close to  $\omega$.

The formulas we have derived for the diffusion coefficient, Eqs.~\eqref{DJ} and \eqref{DRW}, are found to be identical to those proposed by Jeans long ago through a different pathway. While Jeans' arguments were insightful and demonstrated a mastery of free path arguments, his deduction of the corrective terms was somewhat hand-waving and lacked a clear identification of their nature ---specifically, that these corrective terms are  $2 \la \rb_k\cdot\rb_{1+k}\ra/\la r_1^2\ra$.  In contrast, our approach not only identifies the nature of the corrective terms but also allows us to quantify the validity of the approximations, at least for the first few terms. We have verified this both numerically and through DSMC simulations for the particular case of hard spheres.

For a dilute gas of hard spheres, we have compared the theoretical predictions of the diffusion coefficient obtained from the random walk and kinetic theory (first and second Sonine approximations  to the Chapman--Enskog solution to the Boltzmann equation) approaches with those obtained by numerically solving the Boltzmann equation by means of the DSMC method \cite{B94}.
We have found that the results provided by the improved random walk formula were quite good, surpassing even the  first Sonine approximation for intruders lighter than the gas molecules.   Furthermore, the improved random walk results closely approximate the second Sonine coefficient for a wide range of molecule mass values.

Finally, we note that the generalization of the present random walk  approach to  mixtures with arbitrary concentration and/or other type of interactions and dimensions can be easily performed. As a different project, we also plan to extend the present results to the case of granular gases modeled as a gas of smooth hard spheres with inelastic collisions \cite{GarzoBook19}.
Work along these lines is currently in progress.

\acknowledgments

We acknowledge financial support from Grant PID2020-112936GB-I00 funded by MCIN/AEI/10.13039/501100011033, and  from Grant IB20079 funded by Junta de Extremadura (Spain) and by ERDF A way of making Europe.

\appendix
\section{An alternative deduction of Eq.~\eqref{rbrkok} }
\label{ap:alter}
Here, we present an alternative approach to derive Eq.~\eqref{rbrkok} in   Sec.~\ref{sec:3}.  We begin by  assuming  that the sizes of the steps 1 and $k$, as well as their angles, are uncorrelated, so that
  \begin{equation}
  \label{r1rkbis}
\langle \rb_1\cdot \rb_{k}\rangle\approx \langle  {r}_1\rangle  \langle {r}_{k}\rangle \langle \cos \theta_{1,k}\rangle=\lambda^2 \langle \cos \theta_{1,k}\rangle.
  \end{equation}
Next, using the spherical cosine law Eq.~\eqref{cos1k} and considering that $\varphi_{1,2,k}$ is equiprobable, we find
$
\langle \cos\theta_{1k}\rangle\approx\langle\cos\theta_{1,2} \cos\theta_{2,k}\rangle
$.
Now we assume that the dispersion angles of $\theta_{1,2}$ and $\theta_{2,k}$ for two successive collisions are uncorrelated (or weakly correlated), so that
$
\langle \cos_{1k}\rangle\approx\langle\cos_{1,2}\rangle\,\langle\cos\theta_{2,k}\rangle
$.
We estimate  $\la\cos\theta_{2,k}\ra$ as we did with $\la\cos\theta_{1,k}\ra$, and so on. The final result is
\begin{equation}
\label{xk}
\langle \cos_{1k}\rangle = \prod_{i=1}^{k-1}   \la   \cos\theta_{i,i+1} \ra.
\end{equation}
At equilibrium, the value of $i$ is not relevant and  this equation becomes $\langle \cos\theta_{1k}\rangle =\la\cos\theta_{1,2}\ra^{k-1}$. Now we use Eq.~\eqref{rrome} and take the approximation  $\omega\approx\langle v_2/v_1 \rangle \langle \cos\theta\rangle$. We know that $\la v_2\ra=\la v_1 \ra$. This prompts us to make the approximation  $\langle v_2/v_1 \rangle\approx 1$. Thus, $\la\cos\theta_{1,2}\ra\approx \omega$  and Eq.~\eqref{r1rkbis} becomes  Eq.~\eqref{rbrkok}.

\section{Some basic quantities for hard spheres}
\label{ap:basic}

In this appendix, for reference, we provide results on some basic quantities (mean collision time, mean and variance of the free path, free path distribution, and mean persistence ratio)  that appear in the random walk approach  for the particular case of a gas of hard spheres at equilibrium. \sbyr{Recall that} in this case the distribution of velocities $f(\vb)$ is the \sbyr{Maxwell--Boltzmann} distribution \eqref{fMaxwell}.

To evaluate averages with respect to the trajectory of a molecule, we need to know the pdf $P_N$. From Eq.~\eqref{PN1N} we see that this requires knowing both the collision rate $\mu$ and the transition rate $W$.
For hard spheres, the collision rate is given by \cite{Yang1949,CC70}
\begin{equation}
\label{mueHS}
\mu(\vb)=n_2 \sigma_{12}^2\; \left(\frac{2\pi k_\text{B} T}{m_2}\right)^{1/2}\, E(y)
\end{equation}
where
\begin{equation}
\label{ydef}
\yb=\left(\frac{m_2}{2 k_\text{B} T}\right)^{1/2}\; \vb,
\end{equation}
and \footnote{Be aware that this function  $E$ is not  the function with this name in \cite{CC70} (see Eq.~5.4.9 there). The $E$  function in \cite{CC70} is $y$ times the function $E$ defined here.}
\begin{equation}
\label{Edef}
E(y)=e^{-y^2}+\left(2y+\frac{1}{y}\right)\frac{\sqrt{\pi}}{2}\text{Erf}(y).
\end{equation}
Here, $T$ is the temperature of the gas,  $n_2$ is the number density of molecules of the gas, $\sigma_{12}=(\sigma_1+\sigma_2)/2$, where $\sigma_1$ and $\sigma_2$ are the diameters of the intruder and the molecules of the gas, respectively, and Erf is the error function.

Formula \eqref{mueHS} and those we will give hereafter are valid for dilute gases, i.e., for very small $n_2$.  One can extend the validity of these formulas to larger values of $n_2$ by simply realizing that the local density of molecules with which the intruder collides is $g(\sigma_{12})n_2$, with $g(r)$ being the radial distribution function of the gas.  This implies substituting $n_2$ for $g(\sigma_{12})n_2$ in all expressions where $n_2$ appears (which we will not do in this article). Precise values of the radial distribution function at contact, $g(\sigma_{12})$, for mixtures of elastic hard spheres are well known (see Ref.~\cite{Santos2020}, for example).

The transition rate $W(\vb_1,\vb_2)$ for hard spheres is \sbyr{\cite{Yang1949,Visco2008}}  
\begin{equation}
\label{WeHS}
W(\vb_1,\vb_2) =  \frac{n_2 \sigma_{12}^2}{\sqrt{\pi}} \left(\frac{m_2}{2 k_\text{B} T}\right)  \left(\frac{m_1+m_2}{2 m_2}\right)^{2}\; \widetilde{W}(\yb_1,\yb_2)
\end{equation}
with
\begin{align}
\label{WeHSt}
\widetilde{W}(\yb_1,\yb_2)&= \frac{\exp\left\{ - \left[\frac{m_1+m_2}{2 m_2}|\yb_2-\yb_1|+\yb_1\cdot\frac{\yb_2-\yb_1}{|\yb_2-\yb_1|}\right]^2\right\}}{|\yb_2-\yb_1|} .
\end{align}

From Eq.~\eqref{mueHS} one finds the well-known values of the average collision rate
\begin{align}
\label{xaa}
\langle \mu \rangle=& \int d\vb\, f(\vb)  \mu(\vb)  \nonumber\\
=&2 \sqrt{\pi}\,\left(\frac{m_1+m_2}{m_1}\right)^{1/2}\left(\frac{2k_\text{B} T}{m_2}\right)^{1/2} n_2\sigma_{12}^2,
\end{align}
and the mean free path
\begin{align}
\label{xbb}
\lambda\equiv \langle r_1\rangle=&  \int_0^\infty d\tau_1  \int d\vb_1 \,\tilde P_1(\vb_1) (v_1\tau_{1})\nonumber\\
=& \frac{1}{\pi \left(1+m_1/m_2\right)^{1/2}\, n_2\sigma_{12}^2}.
\end{align}
Here, the relation  $r_1=v_1\tau_{1}$ has been used. From these results and taking into account Eq.\ \eqref{0.1} and that $\bar{v}=\lambda \la \mu\ra=(8k_\text{B} T/\pi m_1)^{1/2}$, one gets
\begin{equation}
\label{DKT0eHS}
D_\text{eKT}=\frac{1}{3\pi \, \left(1+m_1/m_2\right)^{1/2}\,n_2 \sigma_{12}^2}   \sqrt{\frac{8k_\text{B} T}{\pi m_1}}
\end{equation}
for hard spheres.

With the expression~\eqref{WeHS} at hand, one can evaluate the mean persistence ratio as
\begin{equation}
\label{xf}
\omega=\int \tilde P_2(1,2) \frac{v_2}{v_1}\cos\theta_{12} d\vb_1  d\vb_2 d\tau_{1}d\tau_{2},
\end{equation}
or equivalently, as
\begin{align}
\label{xomega}
\omega =&\left\langle \frac{v_2}{v_1}\cos\theta_{12}\right\rangle\nonumber\\
=&\frac{1}{   \langle\mu \rangle} \int f(\vb_1)  W(\vb_1,\vb_2)  \frac{v_2}{v_1}\cos\theta_{12}  d\vb_1  d\vb_2.
\end{align}
The integral \eqref{xomega} can be evaluated explicitly (see Eq. (5.51,2) of Ref.\ \cite{CC70}) and the result is
\begin{equation}
\label{omega}
 \omega = \frac{m_1/m_2+ \widetilde{\omega}(m_1/m_2)}{1+m_1/m_2},
\end{equation}
where
\begin{equation}
\label{omegat}
 \widetilde{\omega}(m)=\frac{m}{2}\left[ \frac{m}{\sqrt{1+m}} \ln\left( \frac{1+\sqrt{1+m}}{\sqrt{m}} \right) -1\right].
\end{equation}
Figure \ref{fig:ck} shows the mean persistence ratio $\omega$ (apart from the ratios $c_k/c_{k-1}$) as a function of the mass ratio $m_1/m_2$.
We see that $\omega$ is smaller than unity unless $m_1/m_2\to\infty$. This  implies that the approximation Eq.~\eqref{sup2} is excellent, except when $m_1/m_2$ becomes very large. The fact that $\omega\to 1$  when $m_1\gg m_2$  makes sense since when the intruder is heavier than the particles of the gas,   it retains more of its trajectory and velocity after collisions.  Conversely,  if  $m_1/m_2\to 0$, then we see that $\omega\to 0$. This is  because the lighter the intruder, the less capacity it has to keep  its trajectory and velocity after collisions.

\section{\sbyr{Asymptotic expressions  of the pdf $P(\ell)$ of the free path}}
\label{ap:Asymptotic}
\sbyr{
In this appendix we provide the asymptotic expressions for the pdf $P(\ell)$ given in Eq.~\eqref{xd} when $m\to 0$ and/or $m\to \infty$ (for $\ell$ finite) and when $\ell \to \infty$ (for $m$ finite).

The expression of $P(\ell)$ in the  case of a light intruder moving in a sea of very massive molecules can be obtained from the asymptotic evaluation of the integral $I_{\ell}(x,m)$  of Eq.~\eqref{xe} for $m\to 0$. In this limiting case, only large values of $y$ are relevant in the integrand of $I_{\ell}(x,m)$. Thus, when $y\to\infty$, $E(y)\sim \sqrt{\pi}\, y$ and $I_{\ell}(x,m\to 0)$ behaves as
\begin{align}
I_{\ell}(x,m\to 0)&\sim \pi\,e^{-x} \int_0^\infty y^3 e^{-my^2} dy \nonumber \\
&\sim \frac{\pi}{2m^2}\,e^{-x}.
\end{align}
Introducing this result into Eq.~\eqref{xd}, one  gets
\begin{equation}
\lambda P(\ell)\sim \exp(-\ell/\lambda)
\end{equation}
for $m\to 0$.
}

\sbyr{An asymptotic expression in the opposite limit of a very massive intruder moving in a sea of very light particles ($m\to \infty$) can also be obtained when one realizes that for $m\to\infty$ only small values of $y$ are relevant in the integrand of $I_{\ell}(x,m)$. Taking into account that $E(y)\sim 2$ for   $y\to 0$, one finds
\beqa
\label{v2}
I_{\ell}(x,m)&\sim& 4\int_0^\infty y\exp\left(-my^2-\frac{2x}{y\sqrt{\pi (1+m)}}\right) dy \nonumber\\
&\sim&\frac{2x^2}{(1+m)\pi^{3/2}}\displaystyle G_{0,3}^{\,3,0}\!\left(\left.{\begin{matrix}
 \,\\-1,-\frac{1}{2},0
 \end{matrix}}\;\right|\,\frac{mx^2}{(1+m)\pi}\right),
 \nonumber\\
 \eeqa
where $G_{0,3}^{\,3,0}$ is a Meijer G-function. Substituting  Eq.~\eqref{v2} into Eq.~\eqref{xd}, and taking the limit $m\to \infty$ in the resulting expression, one finds
\begin{equation}
\label{v3}
\lambda P(\ell)
\sim
\frac{4}{\pi^{3/2}} \displaystyle G_{0,3}^{3,0}\!\left(\left.\begin{array}{c}
\,\\ 0,\frac{1}{2},1
\end{array}\;\right|\,\frac{\ell^2}{\pi\lambda^2}\right).
\end{equation}

From the behavior of the Meijer G-function  $G_{0,3}^{3,0}$ for small and large arguments \cite{WolframMeijerG}, one gets
\begin{equation}
\lambda P(\ell)\sim \frac{4}{\pi}
\end{equation}
for small $\ell/\lambda$, and
\begin{equation}
\lambda P(\ell) \sim \frac{8}{\sqrt{3} \pi^{2/3}} \left(\frac{\ell}{\lambda}\right)^{1/3}\exp\left(-\frac{3}{\pi^{1/3}} \left(\frac{\ell}{\lambda}\right)^{2/3}\right)
\end{equation}
for large $\ell/\lambda$.

Finally, for  $\ell/\lambda\to \infty$ but finite $m$,  only large values of $y$ are relevant in the integrand of $I_\ell (x,m)$. In this case,
\begin{align}
I_{\ell}(x,m)&\sim \pi e^{-x/\sqrt{1+m}}\int_0^\infty y^3\, e^{-m y^2} dy \nonumber\\
&\sim \frac{\pi}{2m^2}\,  e^{-x/\sqrt{1+m}}
\end{align}
and hence, $\lambda P(\ell)$ behaves as
\begin{equation}
\lambda P(\ell) \sim \frac{1}{1+m}  \exp\left(- \frac{\ell/\lambda}{\sqrt{1+m}} \right).
\end{equation}
For $m=1$ this expression reduces to that obtained in Ref.~\cite{Visco2008}. When $m=0$, the asymptotic form of $P(\ell)$ for $m\to 0$ is recovered.}

\section{Direct Simulation Monte Carlo (DSMC) method}
\label{ap:dsmc}

The direct simulation Monte Carlo (DSMC) method is employed in this paper to assess the various approximations made for calculating both the coefficients of the reduced collisional series $c_k$ and the diffusion coefficient $D_\text{RW}$.

In the context of a rarefied gas regime dominated by collisions, the DSMC method emerges as an alternative yet complementary approach for solving the Boltzmann equation. Originally introduced by Bird \cite{B63}, classical DSMC simulations were specifically designed to address rarefied gas flows that are computationally inaccessible via molecular dynamics simulations \cite{B70b, B94}. In this work, we follow similar steps as those proposed by P{\"o}schel and Schwager \cite{PS05} to solve the Boltzmann equation for a mixture composed by hard spheres.

The simulation is initiated by drawing the particle velocities from the corresponding Maxwell distribution $f_i$ [see Eq.\ \eqref{fMaxwell}] at an initial temperature $T_{0}$ for each species $i$. The discretized distribution function $f_i^{(\mathcal{N}_i)}$ of species $i$ is derived from  the velocities $\left\{\mathbf{v}_k\right\}$ of $\mathcal{N}_i$ ``virtual particles'' as given by:
\beq\label{AP21}
f_i^{(\mathcal{N}_i)}(\mathbf{v};t)\to \frac{n_i}{N_i}\sum_{k=1}^{\mathcal{N}_i}\delta[\mathbf{v}-\mathbf{v}_k(t)],
\eeq
where $\delta$ is the Dirac delta function.
For low-density regimes, since the collisions are assumed to be instantaneous, the free flight of particles and the collision stage can be temporally decoupled. The DSMC method maintains this assumption, leading to a division into two distinct steps: the convective and collision stages. However, since our simulations assume equilibrium for the molecular gas, only the collision stage is detailed here. The ``general'' procedure can be summarized as follows:
\begin{enumerate}
    \item  To simulate collisions between particles of species $i$ and $j$, a required number of $N_{ij}^{\Delta t}$ candidate pairs for collision in a time $\Delta t$ is selected. This number is determined by \cite{PS05}:
    \beq\label{AP22}
    \mathcal{N}_{ij}^{\Delta t}=\pi \mathcal{N}_in_j\sigma_{ij}^2 g_{ij}^\text{max}\Delta t,
    \eeq
     where $\mathcal{N}_i$ is the total number of simulated particles of species $i$, and $g_{ij}^\text{max}$ is an upper bound for the average relative velocity between two particles. A suitable estimate is  $g_{ij}^\text{max}=Cv^\text{th}_{ij}$, where $v^\text{th}_{ij}=\sqrt{2T_{0}/\overline{m}}$ is the mean thermal velocity, $\overline{m}=(m_i+m_j)/2$, and $C$ is a constant, e.g., $C = 5$ \cite{B94}.

    \item A colliding direction $\widehat{{\boldsymbol {\sigma }}}_{k\ell}$ for a pair of colliding particles labeled as $k$ and $\ell$ is randomly selected with equal probability.

    \item The collision is accepted if
    \beq
    \label{AP23}
    |\widehat{{\boldsymbol {\sigma }}}_{k\ell}\cdot \mathbf{g}_{k\ell}|\equiv |\widehat{{\boldsymbol {\sigma }}}_{k\ell}\cdot (\mathbf{v}_{k}-\mathbf{v}_\ell)|>\mathcal{R}(0,1)g_{ij}^\text{max},
    \eeq
   where $\mathcal{R}(0,1)$ is a uniformly distributed random number in the interval $[0,1]$.

    \item If the collision is accepted, then the velocities of particles are updated according to the scattering rules \cite{GarzoBook19}:
    \beqa
\label{AP24}
& &\mathbf{v}_k\to\mathbf{v}_k-2\mu_{ji}({\mathbf g}_{k\ell}\cdot\widehat{\boldsymbol{\sigma}}_{k\ell})\widehat{\boldsymbol{\sigma}}_{k\ell},\nonumber\\
& &\mathbf{v}_\ell\to\mathbf{v}_\ell+2\mu_{ij}({\mathbf g}_{k\ell}\cdot\widehat{\boldsymbol{\sigma}}_{k\ell})\widehat{\boldsymbol{\sigma}}_{k\ell},
\eeqa
being $\mu_{ij}=m_i/(m_i+m_j)$.
    \item Repeat the procedure for all the permutations of $i = 1$, $2$ and $j = 1$, $2$.
\end{enumerate}

Since our interest lies in studying a mixture where one species is present in tracer concentration, certain modifications are made to the algorithm. Assuming $i\equiv 1$ (intruders) and $j\equiv 2$ (molecular gas), collisions 1-1 are neglected, and when a collision 1-2 or 2-1 occurs, only the velocity of the intruder is modified according to the scattering rules given in step 4. Although the molecular gas is inherently in equilibrium, described by the Maxwell distribution, it is crucial to consider collisions between molecules (labeled as 2-2) in the algorithm. This consideration helps prevent correlations stemming from a distribution biased by the initial velocities assigned to the finite number of molecules $\mathcal{N}_2$.


We measure the diffusion coefficient $D$  of an intruder immersed in a dilute gas at equilibrium by employing the Einstein relation:
\beq
\label{5.1}
D=\frac{1}{6 \Delta t}\left[\langle R^2(t+\Delta t)\rangle-\langle  R^2(t)\rangle\right].
\eeq
Here, $\langle \cdots \rangle$ denotes the average over the $\mathcal{N}_1$ intruders.

\subsection{Coefficients of the reduced collisional series $c_k$}
To measure the projection over a free path of the subsequent $n$-th free path, we must track the trajectory of the intruders between collisions
to find $\rb_n$.
 What one would be tempted to do would be to perform averaging of $\rb_1\cdot\rb_k$ over the subset of realizations where the algorithm detects the very first collision of the intruder with a molecule of the gas.  However, this first accepted collision is likely to be one in which the relative velocity between the colliding particles is quite large, biasing the distribution function of the particles involved in the averaged quantities [see Eq.\ \eqref{AP23}].
Therefore, to avoid that, the coefficients $c_k$ are computed by considering the position of the intruder after every collision $n$ by
\beq
c_k=2\frac{\langle \rb_n\cdot\rb_{n+k}\rangle}{\langle r_n^2\rangle}.
\eeq
This ensures that $f_i$ is the correct one.




\begin{thebibliography}{26}%
\makeatletter
\providecommand \@ifxundefined [1]{%
 \@ifx{#1\undefined}
}%
\providecommand \@ifnum [1]{%
 \ifnum #1\expandafter \@firstoftwo
 \else \expandafter \@secondoftwo
 \fi
}%
\providecommand \@ifx [1]{%
 \ifx #1\expandafter \@firstoftwo
 \else \expandafter \@secondoftwo
 \fi
}%
\providecommand \natexlab [1]{#1}%
\providecommand \enquote  [1]{``#1''}%
\providecommand \bibnamefont  [1]{#1}%
\providecommand \bibfnamefont [1]{#1}%
\providecommand \citenamefont [1]{#1}%
\providecommand \href@noop [0]{\@secondoftwo}%
\providecommand \href [0]{\begingroup \@sanitize@url \@href}%
\providecommand \@href[1]{\@@startlink{#1}\@@href}%
\providecommand \@@href[1]{\endgroup#1\@@endlink}%
\providecommand \@sanitize@url [0]{\catcode `\\12\catcode `\$12\catcode
  `\&12\catcode `\#12\catcode `\^12\catcode `\_12\catcode `\%12\relax}%
\providecommand \@@startlink[1]{}%
\providecommand \@@endlink[0]{}%
\providecommand \url  [0]{\begingroup\@sanitize@url \@url }%
\providecommand \@url [1]{\endgroup\@href {#1}{\urlprefix }}%
\providecommand \urlprefix  [0]{URL }%
\providecommand \Eprint [0]{\href }%
\providecommand \doibase [0]{https://doi.org/}%
\providecommand \selectlanguage [0]{\@gobble}%
\providecommand \bibinfo  [0]{\@secondoftwo}%
\providecommand \bibfield  [0]{\@secondoftwo}%
\providecommand \translation [1]{[#1]}%
\providecommand \BibitemOpen [0]{}%
\providecommand \bibitemStop [0]{}%
\providecommand \bibitemNoStop [0]{.\EOS\space}%
\providecommand \EOS [0]{\spacefactor3000\relax}%
\providecommand \BibitemShut  [1]{\csname bibitem#1\endcsname}%
\let\auto@bib@innerbib\@empty
\bibitem [{\citenamefont {Chapman}\ and\ \citenamefont {Cowling}(1970)}]{CC70}%
  \BibitemOpen
  \bibfield  {author} {\bibinfo {author} {\bibfnamefont {S.}~\bibnamefont
  {Chapman}}\ and\ \bibinfo {author} {\bibfnamefont {T.~G.}\ \bibnamefont
  {Cowling}},\ }\href@noop {} {\emph {\bibinfo {title} {The Mathematical Theory
  of Nonuniform Gases}}}\ (\bibinfo  {publisher} {Cambridge University Press,
  Cambridge},\ \bibinfo {year} {1970})\BibitemShut {NoStop}%
\bibitem [{\citenamefont {Cercignani}(1998)}]{C98}%
  \BibitemOpen
  \bibfield  {author} {\bibinfo {author} {\bibfnamefont {C.}~\bibnamefont
  {Cercignani}},\ }\href@noop {} {\emph {\bibinfo {title} {Ludwig Boltzmann.
  The {M}an {W}ho {T}rusted {A}toms}}}\ (\bibinfo  {publisher} {Oxford
  University Press},\ \bibinfo {year} {1998})\BibitemShut {NoStop}%
\bibitem [{\citenamefont {Reif}(1965)}]{Reif1965}%
  \BibitemOpen
  \bibfield  {author} {\bibinfo {author} {\bibfnamefont {F.}~\bibnamefont
  {Reif}},\ }\href@noop {} {\emph {\bibinfo {title} {Fundamentals of
  {S}tatistical and {T}hermal {P}hysics}}},\ \bibinfo {edition} {international
  student}\ ed.\ (\bibinfo  {publisher} {McGraw-Hill Kogakusha Tokyo},\
  \bibinfo {year} {1965})\BibitemShut {NoStop}%
\bibitem [{\citenamefont {McQuarrie}(1975)}]{McQuarrie1976}%
  \BibitemOpen
  \bibfield  {author} {\bibinfo {author} {\bibfnamefont {D.~A.}\ \bibnamefont
  {McQuarrie}},\ }\href@noop {} {\emph {\bibinfo {title} {Statistical
  {M}echanics}}}\ (\bibinfo  {publisher} {Harper and Row},\ \bibinfo {year}
  {1975})\BibitemShut {NoStop}%
\bibitem [{\citenamefont {Furry}(1948)}]{Furry1948}%
  \BibitemOpen
  \bibfield  {author} {\bibinfo {author} {\bibfnamefont {W.~H.}\ \bibnamefont
  {Furry}},\ }\bibfield  {title} {\bibinfo {title} {On the elementary
  explanation of diffusion phenomena in gases},\ }\href
  {https://doi.org/10.1119/1.1991051} {\bibfield  {journal} {\bibinfo
  {journal} {Am. J. Phys.}\ }\textbf {\bibinfo {volume} {16}},\ \bibinfo
  {pages} {63} (\bibinfo {year} {1948})}\BibitemShut {NoStop}%
\bibitem [{\citenamefont {Yang}(1949)}]{Yang1949}%
  \BibitemOpen
  \bibfield  {author} {\bibinfo {author} {\bibfnamefont {L.~M.}\ \bibnamefont
  {Yang}},\ }\bibfield  {title} {\bibinfo {title} {Kinetic theory of diffusion
  in gases and liquids {I}. {D}iffusion and the {B}rownian motion},\ }\href
  {http://doi.org/10.1098/rspa.1949.0089} {\bibfield  {journal} {\bibinfo
  {journal} {Proc. R. Soc. Lond. A}\ }\textbf {\bibinfo {volume} {198}},\
  \bibinfo {pages} {94} (\bibinfo {year} {1949})}\BibitemShut {NoStop}%
\bibitem [{\citenamefont {Furry}\ and\ \citenamefont
  {Pitkanen}(1951)}]{Furry1951}%
  \BibitemOpen
  \bibfield  {author} {\bibinfo {author} {\bibfnamefont {W.~H.}\ \bibnamefont
  {Furry}}\ and\ \bibinfo {author} {\bibfnamefont {P.~H.}\ \bibnamefont
  {Pitkanen}},\ }\bibfield  {title} {\bibinfo {title} {Gaseous diffusion as a
  random process},\ }\href {https://doi.org/10.1063/1.1748342} {\bibfield
  {journal} {\bibinfo  {journal} {J. Chem. Phys.}\ }\textbf {\bibinfo {volume}
  {19}},\ \bibinfo {pages} {729} (\bibinfo {year} {1951})}\BibitemShut
  {NoStop}%
\bibitem [{\citenamefont {Monchick}(1962)}]{Monchick1962}%
  \BibitemOpen
  \bibfield  {author} {\bibinfo {author} {\bibfnamefont {L.}~\bibnamefont
  {Monchick}},\ }\bibfield  {title} {\bibinfo {title} {Equivalence of the
  {C}hapman-{E}nskog and the mean-free-path theory of gases},\ }\href
  {https://doi.org/10.1063/1.1706535} {\bibfield  {journal} {\bibinfo
  {journal} {Phys. Fluids}\ }\textbf {\bibinfo {volume} {5}},\ \bibinfo {pages}
  {1393} (\bibinfo {year} {1962})}\BibitemShut {NoStop}%
\bibitem [{\citenamefont {{Kosov}}(1982)}]{Kosov1982}%
  \BibitemOpen
  \bibfield  {author} {\bibinfo {author} {\bibfnamefont {N.~D.}\ \bibnamefont
  {{Kosov}}},\ }\bibfield  {title} {\bibinfo {title} {{Elementary kinetic
  theory of diffusion in gases}},\ }\href {https://doi.org/10.1007/BF00827267}
  {\bibfield  {journal} {\bibinfo  {journal} {J. Eng. Phys. Thermophys.}\
  }\textbf {\bibinfo {volume} {42}},\ \bibinfo {pages} {181} (\bibinfo {year}
  {1982})}\BibitemShut {NoStop}%
\bibitem [{\citenamefont {Jeans}(1904)}]{Jeans1904}%
  \BibitemOpen
  \bibfield  {author} {\bibinfo {author} {\bibfnamefont {J.}~\bibnamefont
  {Jeans}},\ }\bibfield  {title} {\bibinfo {title} {L{XX}. {T}he persistence of
  molecular velocities in the kinetic theory of gases},\ }\href
  {https://doi.org/10.1080/14786440409463242} {\bibfield  {journal} {\bibinfo
  {journal} {London Edinburgh Philos. Mag. J. Sci.}\ }\textbf {\bibinfo
  {volume} {8}},\ \bibinfo {pages} {700} (\bibinfo {year} {1904})}\BibitemShut
  {NoStop}%
\bibitem [{\citenamefont {Jeans}(2009)}]{Jeans1940}%
  \BibitemOpen
  \bibfield  {author} {\bibinfo {author} {\bibfnamefont {J.}~\bibnamefont
  {Jeans}},\ }\href@noop {} {\emph {\bibinfo {title} {An Introduction to the
  Kinetic Theory of Gases}}}\ (\bibinfo  {publisher} {Cambridge University
  Press},\ \bibinfo {year} {2009})\BibitemShut {NoStop}%
\bibitem [{\citenamefont {Monchick}\ and\ \citenamefont
  {Mason}(1967)}]{Monchick1967}%
  \BibitemOpen
  \bibfield  {author} {\bibinfo {author} {\bibfnamefont {L.}~\bibnamefont
  {Monchick}}\ and\ \bibinfo {author} {\bibfnamefont {E.~A.}\ \bibnamefont
  {Mason}},\ }\bibfield  {title} {\bibinfo {title} {{Free-Flight Theory of Gas
  Mixtures}},\ }\href {https://doi.org/10.1063/1.1762296} {\bibfield  {journal}
  {\bibinfo  {journal} {Phys. Fluids}\ }\textbf {\bibinfo {volume} {10}},\
  \bibinfo {pages} {1377} (\bibinfo {year} {1967})}\BibitemShut {NoStop}%
\bibitem [{\citenamefont {Bird}(1994)}]{B94}%
  \BibitemOpen
  \bibfield  {author} {\bibinfo {author} {\bibfnamefont {G.~A.}\ \bibnamefont
  {Bird}},\ }\href@noop {} {\emph {\bibinfo {title} {Molecular Gas Dynamics and
  the Direct Simulation Monte Carlo of Gas Flows}}}\ (\bibinfo  {publisher}
  {Clarendon, Oxford},\ \bibinfo {year} {1994})\BibitemShut {NoStop}%
\bibitem [{\citenamefont {Garz\'o}(2019)}]{GarzoBook19}%
  \BibitemOpen
  \bibfield  {author} {\bibinfo {author} {\bibfnamefont {V.}~\bibnamefont
  {Garz\'o}},\ }\href@noop {} {\emph {\bibinfo {title} {Granular Gaseous
  Flows}}}\ (\bibinfo  {publisher} {Springer Nature, Cham},\ \bibinfo {year}
  {2019})\BibitemShut {NoStop}%
\bibitem [{\citenamefont {Paik}(2014)}]{Paik2014}%
  \BibitemOpen
  \bibfield  {author} {\bibinfo {author} {\bibfnamefont {S.~T.}\ \bibnamefont
  {Paik}},\ }\bibfield  {title} {\bibinfo {title} {{Is the mean free path the
  mean of a distribution?}},\ }\href {https://doi.org/10.1119/1.4869185}
  {\bibfield  {journal} {\bibinfo  {journal} {Am. J. Phys.}\ }\textbf {\bibinfo
  {volume} {82}},\ \bibinfo {pages} {602} (\bibinfo {year} {2014})}\BibitemShut
  {NoStop}%
\bibitem [{\citenamefont {Bouchaud}\ and\ \citenamefont
  {Georges}(1990)}]{Bouchaud1990}%
  \BibitemOpen
  \bibfield  {author} {\bibinfo {author} {\bibfnamefont {J.-P.}\ \bibnamefont
  {Bouchaud}}\ and\ \bibinfo {author} {\bibfnamefont {A.}~\bibnamefont
  {Georges}},\ }\bibfield  {title} {\bibinfo {title} {Anomalous diffusion in
  disordered media: Statistical mechanisms, models and physical applications},\
  }\href {https://doi.org/https://doi.org/10.1016/0370-1573(90)90099-N}
  {\bibfield  {journal} {\bibinfo  {journal} {Physics Reports}\ }\textbf
  {\bibinfo {volume} {195}},\ \bibinfo {pages} {127} (\bibinfo {year}
  {1990})}\BibitemShut {NoStop}%
\bibitem [{Note1()}]{Note1}%
  \BibitemOpen
  \bibinfo {note} {Note that the meaning of $f(\protect \mathbf {v})$ here
  differs from that in Ref~\cite {Yang1949}: $f(\protect \mathbf {v})$ is what
  is called $f_1(\protect \mathbf {v})/n_1$ in Ref~\cite {Yang1949}, $n_1$
  being the number density of molecule 1}\BibitemShut {NoStop}%
\bibitem [{Note2()}]{Note2}%
  \BibitemOpen
  \bibinfo {note} {To evaluate $\kappa $ for $m_1/m_2\to \infty $ one first
  notes that, due to the term $\exp (-m y^2)$ in the integrand of $I_{1,1}(m)$,
  the only relevant values of $y$ for the evaluation of this integral are those
  around zero. But $\lim _{y\to 0}E(y)=2$ so that the dominant term of
  $I_{1,1}$ is given by half the integral of $y^4 \exp (-m y^2)$. This integral
  is equal to $3\protect \sqrt {\pi }/8m^{5/2}$, and the value of $\kappa =3\pi
  /8$, for $m_1/m_2\to \infty $, follows.}\BibitemShut {Stop}%
\bibitem [{\citenamefont {Visco}\ \emph {et~al.}(2008)\citenamefont {Visco},
  \citenamefont {van Wijland},\ and\ \citenamefont {Trizac}}]{Visco2008}%
  \BibitemOpen
  \bibfield  {author} {\bibinfo {author} {\bibfnamefont {P.}~\bibnamefont
  {Visco}}, \bibinfo {author} {\bibfnamefont {F.}~\bibnamefont {van Wijland}},\
  and\ \bibinfo {author} {\bibfnamefont {E.}~\bibnamefont {Trizac}},\
  }\bibfield  {title} {\bibinfo {title} {Collisional statistics of the
  hard-sphere gas},\ }\href {https://doi.org/10.1103/PhysRevE.77.041117}
  {\bibfield  {journal} {\bibinfo  {journal} {Phys. Rev. E}\ }\textbf {\bibinfo
  {volume} {77}},\ \bibinfo {pages} {041117} (\bibinfo {year}
  {2008})}\BibitemShut {NoStop}%
\bibitem [{\citenamefont {Lue}(2005)}]{Lue2005}%
  \BibitemOpen
  \bibfield  {author} {\bibinfo {author} {\bibfnamefont {L.}~\bibnamefont
  {Lue}},\ }\bibfield  {title} {\bibinfo {title} {{Collision statistics,
  thermodynamics, and transport coefficients of hard hyperspheres in three,
  four, and five dimensions}},\ }\href {https://doi.org/10.1063/1.1834498}
  {\bibfield  {journal} {\bibinfo  {journal} {J. Chem. Phys.}\ }\textbf
  {\bibinfo {volume} {122}},\ \bibinfo {pages} {044513} (\bibinfo {year}
  {2005})}\BibitemShut {NoStop}%
\bibitem [{Wol()}]{WolframMeijerG}%
  \BibitemOpen
  \href@noop {} {\bibinfo {title} {Meijer {G}-{F}unction ({T}he {W}olfram
  {F}unctions {S}ite)}},\ \bibinfo {howpublished}
  {\url{http://functions.wolfram.com/07.34.02.0001.01}}\BibitemShut {NoStop}%
\bibitem [{Note3()}]{Note3}%
  \BibitemOpen
  \bibinfo {note} {Be aware that this function $E$ is not the function with
  this name in \cite {CC70} (see Eq.~5.4.9 there). The $E$ function in \cite
  {CC70} is $y$ times the function $E$ defined here.}\BibitemShut {Stop}%
\bibitem [{\citenamefont {Santos}\ \emph {et~al.}(2020)\citenamefont {Santos},
  \citenamefont {Yuste},\ and\ \citenamefont {L\'opez~de Haro}}]{Santos2020}%
  \BibitemOpen
  \bibfield  {author} {\bibinfo {author} {\bibfnamefont {A.}~\bibnamefont
  {Santos}}, \bibinfo {author} {\bibfnamefont {S.~B.}\ \bibnamefont {Yuste}},\
  and\ \bibinfo {author} {\bibfnamefont {M.}~\bibnamefont {L\'opez~de Haro}},\
  }\bibfield  {title} {\bibinfo {title} {{Structural and thermodynamic
  properties of hard-sphere fluids}},\ }\href
  {https://doi.org/10.1063/5.0023903} {\bibfield  {journal} {\bibinfo
  {journal} {J. Chem. Phys.}\ }\textbf {\bibinfo {volume} {153}},\ \bibinfo
  {pages} {120901} (\bibinfo {year} {2020})}\BibitemShut {NoStop}%
\bibitem [{\citenamefont {Bird}(1963)}]{B63}%
  \BibitemOpen
  \bibfield  {author} {\bibinfo {author} {\bibfnamefont {G.~A.}\ \bibnamefont
  {Bird}},\ }\bibfield  {title} {\bibinfo {title} {Approach to translational
  equilibrium in a rigid sphere gas},\ }\href@noop {} {\bibfield  {journal}
  {\bibinfo  {journal} {Phys. Fluids}\ }\textbf {\bibinfo {volume} {6}},\
  \bibinfo {pages} {1518} (\bibinfo {year} {1963})}\BibitemShut {NoStop}%
\bibitem [{\citenamefont {Bird}(1970)}]{B70b}%
  \BibitemOpen
  \bibfield  {author} {\bibinfo {author} {\bibfnamefont {G.~A.}\ \bibnamefont
  {Bird}},\ }\bibfield  {title} {\bibinfo {title} {Direct simulation and the
  {B}oltzmann equation},\ }\href@noop {} {\bibfield  {journal} {\bibinfo
  {journal} {Phys. Fluids}\ }\textbf {\bibinfo {volume} {13}},\ \bibinfo
  {pages} {2676} (\bibinfo {year} {1970})}\BibitemShut {NoStop}%
\bibitem [{\citenamefont {P{\"o}schel}\ and\ \citenamefont
  {Schwager}(2005)}]{PS05}%
  \BibitemOpen
  \bibfield  {author} {\bibinfo {author} {\bibfnamefont {T.}~\bibnamefont
  {P{\"o}schel}}\ and\ \bibinfo {author} {\bibfnamefont {T.}~\bibnamefont
  {Schwager}},\ }\href@noop {} {\emph {\bibinfo {title} {Computational Granular
  Dynamics: Models and Algorithms}}}\ (\bibinfo  {publisher} {Springer Science
  \& Business Media},\ \bibinfo {year} {2005})\BibitemShut {NoStop}%
\end{thebibliography}%
%

\end{document}